\documentclass[aps,pra,twocolumn,amsmath,amssymb,superscriptaddress]{revtex4-1}

\usepackage{braket, units}
\usepackage{hyperref}
\usepackage{color}
\usepackage{graphicx}
\usepackage{subfigure}

\newcommand{\tr}{\operatorname{tr}}
\newcommand{\rank}{\operatorname{rank}}
\newcommand{\range}{\operatorname{range}}

\newcommand{\knearY}{\ket {\text{near}Y}}
\newcommand{\Znidaric}{\ifmmode \check{Z}\else \v{Z}\fi{}nidari\ifmmode \check{c}\else \v{c}\fi{}}

\usepackage[dvipsnames]{xcolor}
\definecolor{Zcolour}{rgb}{0.992, 0.588, 0.22}
\definecolor{dkgreen}{rgb}{0,0.5,0}
\definecolor{purple}{rgb}{0.5,0,0.5}

\begin{document}
           
\title{Quantum dynamics of thermalizing systems}
\author{Christopher David White}
\affiliation{Institute for Quantum Information and Matter, Caltech}
\email{cdwhite@caltech.edu}

\author{Michael Zaletel}
\affiliation{Department of Physics, Princeton University, Princeton, NJ 08544, USA}

\author{Roger S. K. Mong}
\affiliation{Department of Physics and Astronomy, University of Pittsburgh}

\author{Gil Refael}
\affiliation{Institute for Quantum Information and Matter, Caltech}

\begin{abstract}

We introduce a method ``DMT'' for approximating density operators of 1D systems that, when combined with a standard framework for time evolution (TEBD), makes possible simulation of the dynamics of strongly thermalizing systems to arbitrary times.  We demonstrate that the method performs well for both near-equilibrium initial states (Gibbs states with spatially varying temperatures) and far-from-equilibrium initial states, including quenches across phase transitions and pure states.

\end{abstract}
\maketitle
\tableofcontents

\section{Introduction}
\begin{figure}
  \includegraphics[width=0.33\textwidth]{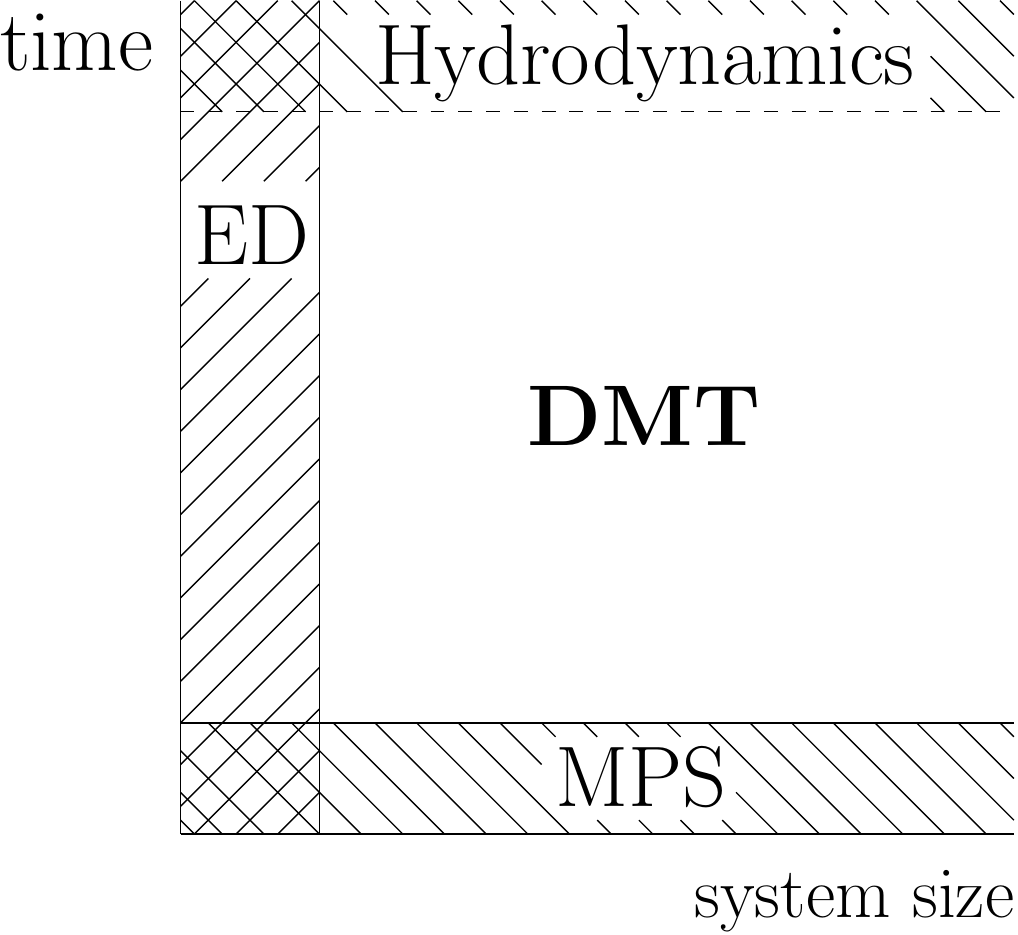}
  \caption{For small systems, exact diagonalization and related methods can treat time evolution of small ETH systems to long times. Matrix product state methods, on the other hand, can treat time evolution of large ETH systems, but only to short times, and hydrodynamic effective theories can phenomenologically describe the long-time limit, but not intermediate times. Our method, DMT, can treat large ETH systems at all times.}
  \label{fig:state_of_field}
\end{figure}

Questions about how (and whether) hydrodynamic behavior emerges from microscopic quantum physics arise frequently in condensed matter physics.
The exploration of this physics is hampered by the limitations of existing numerical methods (cf Fig.~\ref{fig:state_of_field}).
Numerically exact methods (like exact diagonalization and Krylov subspace methods) can treat the dynamical properties of small systems at arbitrary times, but require memory and computation time exponential in system size.
Matrix product state methods, on the other hand, can treat large systems---but only when the systems have little entanglement entropy.
This means that for thermalizing systems, whose entanglement entropy grows linearly with time, MPS methods can only treat short-time behavior (see Figure \ref{fig:state_of_field}).

We introduce a numerical method (``density matrix truncation'' or ``DMT'') based on matrix product representations of density operators.
This algorithm can accurately simulate not only short-time, low-entanglement behavior and long-time, hydrodynamic behavior, but also the complex intermediate-time behavior from which the hydrodynamics emerges.
While using matrix-product representations of mixed states is not new, the core of our algorithm is a new method for truncating matrix product density operators (MPDOs).
This truncation exactly preserves the energy density of the system and other local conserved quantities, with the aim of leaving the hydrodynamics unaffected.
It also avoids generating unphysical density matrices with negative entropy.
At the same time, it is efficient enough that by taking large (but constant-in-system-size) bond dimension one can capture the  thermalization process.

We first (in Sec.~\ref{S:Int}) offer some background on matrix product state methods and intuition for why a method using MPDOs should be able to efficiently simulate time evolution governed by Hamiltonians satisfying the eigenstate thermalization hypothesis (ETH) to arbitrary times. We also motivate certain properties of our method.
We then describe (in Sec.~\ref{S:Method}) our algorithm for time evolution.
This algorithm consists of a standard time-evolution framework very much like that in TEBD (described in Appendix \ref{APP:TEF}), paired with a novel scheme for truncating MPDOs.
We then apply our algorithm to time evolution starting from a pure state (Sec.~\ref{SS:PSE}) and find that it qualitatively improves upon existing methods.
Applied to a mixed state (Sec.~\ref{SS:MSnEE} and App.~\ref{SS:MSffEE}), we find that DMT matches or exceeds the state of the art.
We conclude with directions we hope will improve on the method.

\section{Background and intuition}
\label{S:Int}
\subsection{Background: matrix product state methods}
Simulating real-time evolution of many-body quantum-mechanical systems is hard: a system of $L$ sites generically requires storage and computation time exponential in $L$.
One line of attack, e.g.\ time-evolving block decimation (TEBD) \cite{PhysRevLett.91.147902,PhysRevLett.93.040502, PhysRevLett.93.076401}, proceeds by representing unitary time evolution as a series of small time-steps applied to a matrix-product state (MPS) representation of a pure state.
These matrix-product structures offer efficient representations of certain states (broadly speaking, ``low-entanglement states'') in the sense that typical operations require polynomial time and memory.

MPS simulations of time-evolving pure states are stymied by the fast increase of entanglement entropy with time, which grows linearly in time for a typical global quench.
When one compresses a pure state as an MPS, memory and computation time requirements grow exponentially in the entanglement entropy among the subsystems, so this linear growth in entanglement entropy sets a hard upper limit on the timescales on which MPSs are useful, though a variety of methods have been used in attempts to circumvent this limit.
\cite{1367-2630-8-12-305,PhysRevLett.102.240603,holzner_chebyshev_2011,Muller-HermesCiracBanuls2012,WallCarr-NJP-2012,enss_light_2012,tamascelli_improved_2015,PhysRevB.91.115144,haegeman_time-dependent_2011,haegeman_post-matrix_2013,haegeman_geometry_2014,haegeman_unifying_2016}.
One case in which entanglement growth does not limit the useful timescale of MPS simulations is many-body localized (MBL) dynamics, which exhibits a modest logarithmic entanglement growth \cite{PhysRevLett.109.017202,PhysRevLett.110.260601}.
On the thermal side of the MBL transition, however, entanglement grows quickly, and even the transition itself is expected to show extensive (volume-law) entanglement \cite{2014arXiv1405.1471G}---consequently, pure-state time evolution under Hamiltonians remains restricted to short times.
        
We study time evolution under Hamiltonians that are not many-body localized, for which the long time behavior shows extensive (volume-law) entanglement.
We note a very interesting perspective on this barrier was investigated in the recent work of Leviatan et al. \cite{2017arXiv170208894L}.
They found that by continuing an energy-conserving version of the time evolution despite the large errors associated with staying in the MPS manifold, certain of the statistical aspects of the thermalization process were still correctly captured.

Research into mixed-state time evolution and Lindblad dynamics has also progressed.
It has been proven that density matrices (and purifications) of Gibbs states of local Hamiltonians have efficient matrix product representations \cite{PhysRevLett.93.207205,PhysRevB.73.085115,PhysRevB.91.045138,2014arXiv1410.2224P}.
Two schools of thought have used this insight to develop a series of methods for simulating time evolution.
One school employs density matrices \cite{PhysRevLett.93.207205,PhysRevLett.93.207204,1742-5468-2009-02-P02035,PhysRevB.77.064426,prosen_operator_2007,prosen_quantum_2008,znidaric_complexity_2008,pizorn_operator_2009,PhysRevB.77.064426,benenti_charge_2009,jesenko_finite-temperature_2011,znidaric_solvable_2011,prosen_diffusive_2012,pizorn_real_2014,cui_variational_2015,PhysRevA.92.022116,marzolino_computational_2016}.
They note that the space of operators on a spin chain is the tensor product of onsite operator spaces, just as the space of many-body pure states being a tensor product on onsite Hilbert spaces; the chief difference (in this view) is merely the dimensionality of the onsite space.
For example, on a spin-half chain, the space of onsite operators is four-dimensional, while the space of pure states is two dimensional.
This school then applies familiar pure state methods, including the creation and truncation of matrix product states and time evolution by TEBD, to density matrices---which are, after all, merely vectors in a larger space.
The resulting truncation algorithms minimize the error according to the Hilbert-Schmidt (Frobenius) norm.
In certain situations---in particular, dynamics near thermal equilibrium or a non-equilibrium steady state---this approach works well.
In other situations, however---in particular, time evolution starting from a pure state---the density matrices suffer from a catastrophic loss of positivity.
(Even checking positivity is NP-hard in the system size \cite{PhysRevLett.113.160503}.)

The second school \cite{PhysRevB.79.245101,feiguin_spectral_2010,PhysRevLett.108.227206,enss_light_2012,Barthel-NJP-2013,KBM2013,PhysRevB.90.060406,PhysRevB.92.125119,karrasch_spin_2015,Kennes201637} uses purifications instead of density matrices to represent mixed states.
They pair each site in the system with an ancilla, a notional site representing the bath.
The mixed nature of the system is represented by entanglement between sites and their ancillae, so the system and ancillae together are in a pure state.
Grouping each site and its ancilla into a larger onsite Hilbert space, one can write a matrix product representation for this pure state and apply the usual methods (truncation, TEBD, etc.)
This solves the positivity problem: unlike density matrices, where many operators with reasonable matrix product representations are not positive and hence are invalid as density matrices, every representable vector is a valid state.
Moreover, since one can act with a unitary on the space of ancillae without changing the physical state, one can try to exploit this freedom to reduce the MPS bond dimension of the purification \cite{Barthel-NJP-2013, KBM2013}.
There is also a hybrid approach which locally unzips a density matrix into a purification, which preserves positivity by construction \cite{PhysRevLett.116.237201}.
These purification methods employ truncations which minimize error according to the inner product $\langle \cdot | \cdot \rangle$ defined on the whole (system with ancillae) state.

We argue that neither the Frobenius norm on density matrices nor the quantum-mechanical norm on purifications is the correct notion of error.
In the case of density matrices, the Frobenius norm fails to account for the fact that truncations that change the component of the density-matrix vector along the identity (e.g., which are not trace-preserving) are disastrous, because they can lead to loss of positivity.
Moreover, neither notion of error captures spatial locality: a good notion of error should prioritize short-range properties of the model and guarantee that certain quantities (the local conserved quantities of the model under consideration, like energy density or spin) are unchanged.
Since the methods of both the density-matrix and purification schools generically change the model's conserved quantities at every gate application, they are unable in principle to approach the known ``hydrodynamic'' long-time behavior of systems which thermalize.
This may be the reason that existing density-matrix methods lose accuracy over time, even though one would expect the accuracy of the matrix-product representation to improve as the state approaches equilibrium.

In this work we propose a truncation of density matrices that ameliorates the positivity problem of Frobenius truncation and exactly preserves the expectation values of all operators on all regions of up to three sites in diameter.

\subsection{Intuition: thermalization and computation}
Why should one be able to efficiently simulate the dynamics of a local ETH Hamiltonian?
In the long-time limit, the system is well described (as far as local observables are concerned) by a Gibbs state, which has an efficient matrix product density operator (MPDO) representation \cite{PhysRevB.91.045138,PhysRevB.73.085115}.
Moreover, the system will (one expects) locally thermalize before it reaches global equilibrium, and indeed after some short local thermalization time $t_{\mathrm{therm}}$ expectation values of local operators will be well approximated by the expectation values of those operators in a Gibbs state with spatially varying temperature, chemical potential, etc.
Heuristically, one can imagine keeping the state exactly out to the local thermalization time and then truncating to an efficient representation.
This would require a maximum bond dimension ca. $(d^2)^{vt_{\mathrm{therm}}}$, where $v$ is some entanglement speed and $d$ is the dimension of the onsite Hilbert space.
If $vt_{\mathrm{therm}}$ is not too large, this approach itself may be workable.

In practice, however, one will wish to efficiently represent the state even at early and intermediate times $t < t_{\mathrm{therm}}$---and also to avoid dependence on the hard-to-define and likely-unknown constant $t_{\mathrm{therm}}$.
Having decided upon an MPDO representation, then, one is faced with the problem of writing a truncation algorithm: an algorithm that will approximate a given MPDO by another, more compact MPDO.

The natural approach, by analogy with matrix product states, is to discard low-weight Schmidt vectors. (This approach turns out to be an imperfect solution, but it is a useful first step.) 
We call this truncation ``Frobenius truncation.''
A density operator is a vector in a space with the same tensor-product structure as a state, but a larger onsite dimension.
We can therefore cut the chain at bond $j$ into two sections $L$ and $R$, Schmidt decompose, and truncate it:
\begin{equation}
  \label{eq:ZVtrunc}
  \rho =
  \sum_{\alpha = 0}^{\chi-1} \hat x_{L\alpha} s_\alpha \hat x_{R\alpha}
  \mapsto
  \sum_{\alpha = 0}^{\chi' - 1} \hat x_{L\alpha} s_\alpha \hat x_{R\alpha}, \qquad \chi' < \chi
\end{equation}
where $\hat x_{L\alpha}, \hat x_{R\alpha}$ are operators supported on $L$ and $R$, respectively, and $\tr x_{L\alpha}^\dag x_{L\beta} = \tr x_{R\alpha}^\dag x_{R\beta} = \delta_{\alpha\beta}$.
Explicitly, one starts with a matrix-product representation of the density operator
\begin{equation}
	\rho = \sum_{\alpha = 0}^{\chi-1}
             \big[A_1^{\mu_1} \cdots A_j^{\mu_j}\big]_\alpha
             s_\alpha
             \big[B_{j+1}^{\mu_{j+1}} \cdots B_{L}^{\mu_L}\big]_\alpha
             \hat \sigma_1^{\mu_1}\cdots\hat\sigma_L^{\mu_L},
\end{equation}
where $A_l^{\mu_{l}}, B_l^{\mu_{l}}$ are $\chi\times\chi$ matrices---with the exception of $A_1^{\mu_1}$ and $B_L^{\mu_{L}}$, which are $1\times\chi$ and $\chi\times1$ respectively.
We suppress for compactness a sum on $\mu$ (that is, on a basis for the space on onsite operators, here the Pauli matrices $\sigma^\mu$, with $\sigma^0 = I$). 
The truncation \eqref{eq:ZVtrunc} is then
\begin{align}\begin{split}
	\rho &= \sum_{\alpha = 0}^{\chi - 1}
		\big[ \cdots A_j^{\mu_j} \big]_\alpha
		s_\alpha
		\big[ B_{j+1}^{\mu_{j+1}} \cdots \big]_\alpha
		\hat \sigma_1^{\mu_1}\cdots\hat\sigma_L^{\mu_L}
\\	&\mapsto \sum_{\alpha = 0}^{\chi'-1}
		\big[ \cdots A_j^{\mu_j} \big]_\alpha
		s_\alpha
		\big[ B_{j+1}^{\mu_{j+1}} \cdots \big]_\alpha
		\hat \sigma_1^{\mu_1}\cdots\hat\sigma_L^{\mu_L}, 
\end{split}\end{align}
with $\chi' < \chi$.
This approximation minimizes the Frobenius (Hilbert-Schmidt) norm distance---but a priori that is not the only norm one could use.

The Frobenius approximation scheme gives poor results for initial states far from equilibrium. One can see why by considering the expectation values of the operator $O^y_{lt} = U(t) \sigma^y_l U(t)^\dag$ for a system that starts at $t = 0$ in a product of $\sigma^y$ eigenstates.
(Note that $O^y_{l,t}$ is a Schr\"odinger-picture operator parametrized by time $t$.
We work in the Schr\"odinger picture throughout, except where noted.)
At time $t$, $\langle \psi(t) |  O^{y}_{l,t} |\psi(t)\rangle = \pm 1$---but generically $O^{y}_{l,t}$ will be a large, complicated operator (if we choose $t$ larger than the whole-system entanglement time, as we are free to do, it will have support throughout the system) and essentially unrelated to the local operators we wish to measure.
There are $2^L$ such operators $O^{y}_{l_1l_2\dots,t}  = U(t) \sigma^y_{l_1}\sigma^y_{l_2}\cdots U(-t)$, all corresponding to long-range operators with expectation value $\pm 1$.
These operators $O^{y}_{l_1l_2\dots,t} $ form part of an orthonormal basis for the space of operator space.
Errors along the dimensions $O^{y}_{l_1 l_2 \dots} $ will be penalized by the Frobenius-norm notion of distance with precisely the same severity as errors along more physically relevant dimensions, like $\sigma^y_l$.
A more reasonable metric for truncation error should be willing to ``forget'' this information, in favor of more accurately keeping local operators, once they are no longer expected to feedback into the hydrodynamics.
(There are more worrying problems still with the na\"ive Frobenius truncation, which rapidly leads to a dramatic loss of positivity in the supposed density matrix for many  far-from-equilibrium initial-conditions, but these problems can be remedied by considering purifications.)

A BBGKY-like hierarchy for the dynamics of reduced density matrix of a spin chain offers a clue as to how to proceed.
In a system governed by a Hamiltonian that is the sum of two-site terms $H = \sum_j h_{j,j+1}$, the dynamics of one-site reduced density matrices (say $\rho_{j}$, the reduced density matrix on site $j$) depends on the two site density matrices: 
\begin{align}
  \label{BBGKY1}
  \frac d {dt} \rho_j &= -i \tr_{\{j' \ne j\}} [H, \rho]\notag\\
  &= -i [h_{j, j+1} \rho_{j,j+1}] - i [h_{j-1, j}\rho_{j-1,j}]
\end{align}
where we write $\rho_{j, j+1}$ for the two-site reduced density matrix on sites $j, j+1$. Meanwhile the two-site reduced density matrices depend on three-site density matrices, the three-site on four-site, and so on up the sequence.
One can imagine truncating this hierarchy at some length $l$---that is, tracking $l$-site reduced density matrices for some $l$, say 2 or 3 or 6, and writing the dynamics for the $l$-site density matrices in terms of some approximation for the $l+1$-site density matrices.
A natural choice for such an approximation is to replace the $l+1$-site matrices by their disconnected component, e.g., for $l=1$, take $\rho_{j, j+1} \sim \rho_j \rho_{j+1}$.
The problem of the operators $O^{y}_{l,t}$ then never arises.

The BBGKY-like approach fixes certain problems with Frobenius truncation, but comes with its own set of problems.
It is not obvious that the $l$-site density matrices will even be consistent, in the sense that it may not be possible to write them as reduced density matrices for a density matrix on the whole system. (Checking this, for a given set of reduced density matrices, is QMA-hard \cite{2006quant.ph..4166L}.) 
Moreover, the hard truncation at $l$ sites may not be appropriate to capture the dynamics of the system.
Longer-range operators may feed back into the dynamics of few-site density matrices via  the hierarchy starting with \eqref{BBGKY1}---and conversely, some short-range operators may have a negligible effect on the dynamics of operators of physical interest.

The Frobenius norm attempts to keeps all operators with equal weight, while BBGKY keeps connected components only up to a hard cutoff. 
A natural compromise is to interpolate between the two by weighting the connected components of an operator according to some measure of locality.
In the current work, we take  the first step in this direction: we approximate the whole-system density matrices in such a way that the dynamics of reduced density matrices on up to three sites matches the BBGKY hierarchy, but instead of straightforwardly closing the BBGKY hierarchy at that level, we approximate larger connected components using a method similar in spirit to the Frobenius truncation \eqref{eq:ZVtrunc}.
Our method zeros out long-range correlations, replacing entanglement between different parts of the system by entanglement with a notional bath.

Although we have used thermalizing Hamiltonians to motivate our method, our method does not in fact assume that the Hamiltonian governing the dynamics is thermalizing.
We expect to be able to use it to treat MBL Hamiltonians, with a more stringent accuracy vs. bond dimension tradeoff.

\section{Method: truncation of MPDOs}
\label{S:Method}

\begin{figure}
  \includegraphics[width=0.20\textwidth]{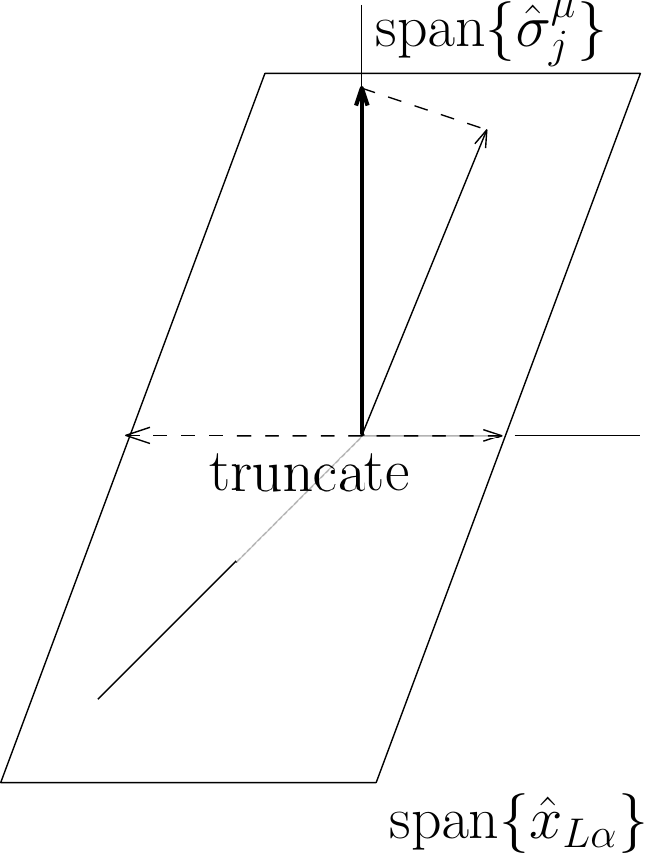}
  \caption{DMT, viewed in the space of operators on the left half of the chain.
    We truncate perpendicular to  certain physically-relevant operators (the $\sigma^\mu_j$).
  }
  \label{fig:methodintuition}
\end{figure}

	Given an MPDO and a particular bond $j$, we wish to truncate the rank of the MPDO. How can we modify the Frobenius truncation \eqref{eq:ZVtrunc} in such a way that it does not change the trace of the density matrix, nor the expectation values of local operators?
The trick is to start by Schmidt decomposing the whole density matrix, and then cleverly choose basis changes on the spaces of left and right Schmidt vectors that put the data we want to avoid changing in specific, easily-understood locations (see Fig.~\ref{fig:methodintuition}).
We can then slot our new truncation into a (slight modification of a) standard time-evolution framework, TEBD (see Appendix \ref{APP:TEF}).

\begin{figure}[b]
  \includegraphics[width=0.45\textwidth]{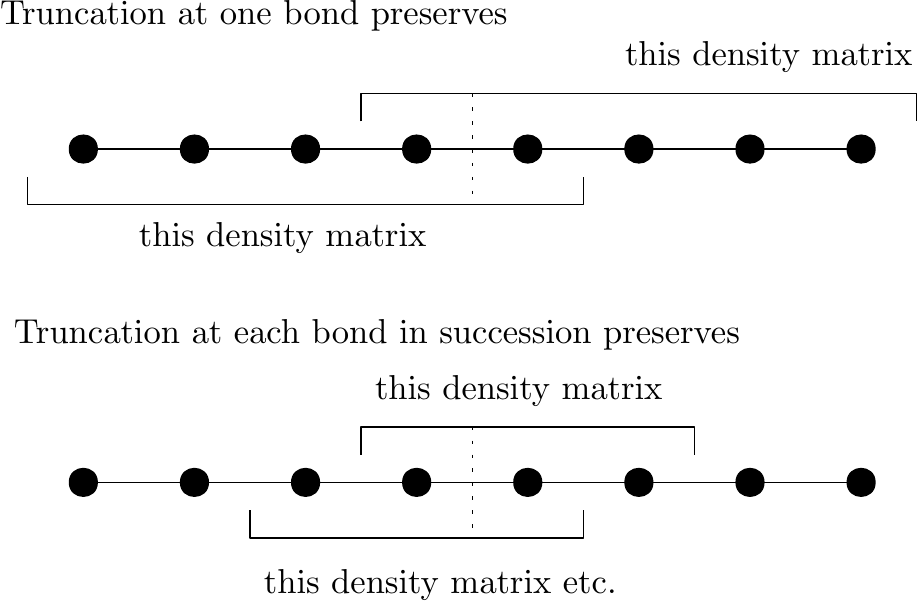}
  \caption{The reduced density matrices that are guaranteed to be preserved under truncation.}
  \label{fig:guarantees}
\end{figure}

Our truncation algorithm guarantees that the following will not change, up to the precision of the numerical linear algebra involved, in a truncation on bond $j$ (cf Fig.~\ref{fig:guarantees}):
\begin{enumerate}
\item the trace of the density matrix, $\tr \rho$;
\item the reduced density matrix $\rho_{1 \cdots j+1}$ on sites $1,\dots,(j+1)$; 
and
\item the reduced density matrix $\rho_{j \cdots L}$ on sites $j,\dots,L$. 
\end{enumerate}
Consequently, no truncation will change the expectation of any operator on three contiguous sites, because any such operator is always contained within one of the guaranteed-preserved density operators.

These guarantees do not fully specify our truncation method.
To do so, define a matrix of connected correlators across the cut $j$:
\begin{equation}
  \tilde M_{\alpha\beta} = \langle \hat y_{L\alpha} \hat y_{R\beta} \rangle - \langle \hat y_{L\alpha}\rangle\langle \hat y_{R\beta} \rangle
\end{equation}
with $\hat y_{L\alpha}$ and $\hat y_{R\beta}$ operators supported on sites $1,\dots,j$ and $j+1,\dots,L$ respectively.
(The $\hat{y}$'s spans the set of observables and are defined below.)
We wish to replace this matrix $\tilde M$ by another $\tilde M'$ with lower rank such that $\tr (\tilde M - \tilde M')^\dag(\tilde M - \tilde M')$ is minimized subject to the constraints above.
The focus on connected components is itself an important improvement.

\subsection{Setting, notation, and tools}
The concept of the method may be straightforward, but it is obscured by a flurry of notation.
We start truncation on bond $j$ with an MPDO of the form 
\begin{align}\begin{split}
	\rho = \sum_{\alpha = 0}^{\chi-1} \sum_{\{\mu\}}
		\big[A_1^{\mu_1} \cdots A_j^{\mu_j}\big]_\alpha
		s_\alpha
		\big[B_{j+1}^{\mu_{j+1}} \cdots B_{L}^{\mu_L}\big]_\alpha
		\\\times \hat\sigma_1^{\mu_1}\cdots\hat\sigma_L^{\mu_L},
\end{split}\end{align}
on an $L$-site chain.
The $A_l^{\mu_{l}}, B_l^{\mu_{l}}$ are $\chi\times\chi$ matrices---with the exception of $A_1^{\mu_1}$ and $B_L^{\mu_{L}}$, which are $1\times\chi$ and $\chi\times1$ respectively.
($\chi$, called the bond dimension, will, in fact, vary between bonds and between steps of time-evolution and truncation, but for the moment we suppress this variation.)

In writing our truncation algorithm, we hat our operators \footnote{We hat all operators except for the density matrix $\rho$, and identity $I$.}.
We use Roman letters (frequently $j,l$) to index sites and bonds;
a bond inherits the index of the site to its left.
We use Greek letters (frequently $\alpha, \beta,\gamma$---but excepting $\mu$ and $\chi$) for the virtual index labeling the Schmidt vectors.
The Greek letter $\mu = 0,1,2,3$ is used to label Pauli matrices, i.e., $\hat\sigma^\mu_j$ is an operator at site $j$ (with $\hat\sigma^0 = I, \hat\sigma^1 = \hat\sigma^x$, etc.).

Following the standard notation \cite{Schollwock201196},
 the MPDO is in mixed-canonical form with an orthogonality center at site $j$---that is, for any $j_1 \le j$ and $j_2 \ge j$, the operators
\begin{align}\begin{split}
	\hat x_{L\alpha}[j_1] &= \sum_{\{\mu\}} \big[A_1^{\mu_1} \cdots A_{j_1}^{\mu_{j_1}}\big]_\alpha\hat\sigma_1^{\mu_1}\cdots\hat\sigma_{j_1}^{\mu_{j_1}} ,
\\	\hat x_{R\alpha}[j_2] &= \sum_{\{\mu\}} \big[B_{j_2+1}^{\mu_{j_2+1}} \cdots B_{L}^{\mu_L}\big]_\alpha\hat \sigma_{j_2+1}^{\mu_{j_2+1}}\cdots\hat\sigma_L^{\mu_L}
\end{split}\end{align}
are orthogonal with respect to the Frobenius inner product
\begin{equation*}
  \tr [x_{L\alpha}^\dag x_{L\beta}] = \tr [x_{R\alpha}^\dag x_{R\beta}] = \delta_{\alpha\beta}.
\end{equation*}
This mixed-canonical form gives the Schmidt decomposition of the density matrix $\rho$ at bond $j$:
\begin{align}
	\rho &= \sum_{\alpha = 0}^{\chi - 1} \hat x_{L\alpha}[j] \ s_\alpha \ \hat x_{R\alpha}[j] .
	\label{eq:rho_split_j}
\end{align}
Henceforth, we will implicitly always be working with Schmidt vectors at bond $j$, and drop the bond label as follows:
\begin{align}\begin{split}
	\hat x_{L\alpha} &= \hat x_{L\alpha}[j] ,
\\	\hat x_{R\alpha} &= \hat x_{R\alpha}[j] .
\end{split}\end{align}
The two vector spaces $\operatorname{span} \{\hat x_{L\alpha}\}$ and $\operatorname{span} \{\hat x_{R\alpha}\}$ are the setting in which we work.

We frequently abuse notation by replacing $s$ for a diagonal matrix whose entries are $s_\alpha$.
This allows us to shorten Eq.~\eqref{eq:rho_split_j} into
\begin{align}
	\rho 
	= \hat x_{L} s \hat x_R,
\end{align}
where $\hat x_{L,R}$ are row and column vectors of operators, respectively.

With stage set and notation defined, we can walk through the algorithm.

\subsection{Re-writing the MPDO to expose properties whose preservation we guarantee}
We wish to take the MPDO
\begin{align}
	\rho = \sum_{\alpha = 0}^{\chi - 1} \hat x_{L\alpha} s_\alpha \hat x_{R\alpha}
\end{align}
(cut along bond $j$) and re-write it as
\begin{align}
  \rho = \sum_{\alpha,\beta = 0}^{\chi - 1} \hat y_{L\alpha} \, M_{\alpha\beta} \, \hat y_{R\beta} ,
\end{align}
with the new bases $\{\hat y_{L\alpha}\}, \{\hat y_{R\alpha}\}$.
The bases $\{\hat y_{L\alpha}\}, \{\hat y_{R\alpha}\}$ and $M$ are chosen such that the properties we wish to avoid changing are characterized by certain easily-identifiable blocks of $M$:
\begin{enumerate}
\item $\tr \rho$ is independent of $M_{\alpha\beta}$ for $\alpha, \beta \ne 0$.
\item The reduced density matrix on sites $1,\dots,(j+1)$, $\rho_{1 \cdots j+1} = \tr_{\{(j+2) \cdots L\}}\rho$, is independent of $M_{\alpha\beta}$ for $\beta \ge 4$.
\item The reduced density matrix on sites $j,\dots,L$, $\rho_{j \cdots L} = \tr_{\{1 \cdots (j-1)\}}\rho$, is independent of $M_{\alpha\beta}$ for $\alpha\ge 4$.
\end{enumerate}
Once we have made this basis change we will be able to modify $M_{\alpha\beta}, \alpha,\beta \ge 4$ with impunity: no such modification will violate our guarantees.

Consider a change of basis
\begin{align}\begin{split}
	\hat y_{L\beta}
		&\equiv (\hat x_{L}Q^*_L)_{\beta}
	\\	&\equiv \sum_{\alpha = 0}^{\chi - 1} \hat x_{L\alpha}Q^*_{L\alpha\beta}
	\\	&= \sum_{\alpha,\{\mu\}} \big[A_1^{\mu_1} \dots A_j^{\mu_j}\big]_{\alpha}Q^*_{L\alpha\beta}
			\,\hat\sigma^{\mu_1}_1 \cdots \hat\sigma_j^{\mu_j} ,
	\\
	\hat y_{R\beta}
		&\equiv (Q^\dag_R\hat x_R)_{\beta}
	\\	&\equiv \sum_{\alpha = 0}^{\chi - 1} \hat Q^*_{R\alpha\beta}x_{R\alpha}
	\\	&= \sum_{\alpha,\{\mu\}} Q^*_{R\alpha\beta}\big[B_{j+1}^{\mu_{j+1}}\dots B_{L}^{\mu_{L}}\big]_\alpha
			\hat\sigma^{\mu_{j+1}}_{j+1} \cdots \hat\sigma_L^{\mu_L} .
\end{split}\end{align}
with $Q_{L,R}$ unitary $\chi\times\chi$ matrices.
Now write
\begin{align}\begin{split}
	\rho &= \hat  x_{L} s \hat x_{R} = [\hat x_{L}Q^*_L][Q_L^T s Q_R][Q_R^\dag \hat x_{R}]
	= \hat y_L M \hat y_R ,
	\label{eq:MPDO_Mdecomp}
\end{split}\end{align}
we can see that $M$ is related to $s$ via
\begin{equation}
	M = Q_L^T s Q_R .
\end{equation}

The requisite basis transformations $Q_{L,R}$ are given by QR decompositions
\begin{align}\begin{split}
	Q_{L\alpha\beta}{R_{L\beta}}^\mu
		&=\tr[\hat x_{L\alpha}\hat\sigma_j^\mu]
	\\	&\propto \big[A_1^{0} \cdots A_{j-1}^0A_j^{\mu}\big]_\alpha \in \mathbf{C}^{\chi\times 4} ,
\\	Q_{R\alpha\beta}{R_{R\beta}}^\mu
		&= \tr[\hat x_{R\alpha}\hat \sigma_{j+1}^\mu]
	\\	&\propto \big[B^\mu_{j+1}\cdots B^{0}_{L-1} B^{0}_L\big]_\alpha \in \mathbf{C}^{\chi\times 4} .
\end{split}\end{align}
(here we use the Einstein summation convention).
In this context, the fact that the fact that the ${R_{L\beta}}^\mu$ is upper triangular is exactly the statement that ${R_{L\beta}}^\mu = 0$ for $\beta > \mu$.
(Similarly for ${R_{R\beta}}^\mu$.)

To see that this is in fact the basis change we seek, first note the trace relations
\begin{align}\begin{split}
	\tr [\hat \sigma^\mu_{j} \hat y_{L\beta} ]
	&= \sum_{\alpha}\tr [\hat\sigma^\mu_j \hat x_{L\alpha}]Q^*_{L\alpha\beta}
	= {R_{L\beta}}^\mu ,
\\	\tr [\hat \sigma^\mu_{j+1} \hat y_{R\beta} ]
	&= \sum_{\alpha} Q^\dag_{R\beta\alpha}\tr [\hat\sigma^\mu_{j+1} \hat x_{L\alpha}]
	= {R_{R\beta}}^\mu .
\end{split}\end{align}
The trace of the density matrix is
\begin{align}\begin{split}
	\tr \rho &= \sum_{\alpha,\beta = 0}^{\chi-1} (\tr \hat y_{L\alpha}) M_{\alpha\beta} (\tr \hat y_{R\beta}) = {R_{L0}}^0M_{00}{R_{R0}}^0 .
\end{split}\end{align}
(Recall that $\hat \sigma^0_{j} = \hat \sigma^0_{j+1} = I$ and that $R$'s are upper triangular.)
This shows that $\tr\rho$ is independent of the majority of $M$, as desired.
Similarly, the density matrices on sites $1,\dots,j+1$ and $j,\dots,L$ are
\begin{subequations}
\begin{align}\begin{split}
	\rho_{1 \cdots j+1}
	&= \sum_{\alpha, \beta = 0}^{\chi - 1} \hat y_{L\alpha} M_{\alpha\beta} 
      \sum_{\mu=0}^3 \frac{ \tr[\hat y_{R\beta} \hat\sigma^\mu_{j+1}] \hat\sigma^\mu_{j+1} }{2}
	\\	&= \frac12 \sum_{\mu=0}^3 \hat\sigma^\mu_{j+1} \sum_{\alpha, \beta = 0}^{\chi - 1}
          \hat y_{L\alpha} M_{\alpha\beta} {R_{R\beta}}^\mu ,
\end{split}\end{align}
and
\begin{align}\begin{split}
	\rho_{j \cdots L}
	&= \sum_{\alpha, \beta = 0}^{\chi - 1}
		\sum_{\mu=0}^3 \frac{ \hat\sigma^\mu_j \tr[\hat\sigma^\mu_j \hat y_{L\alpha}] }{2} M_{\alpha\beta} \hat y_{R\beta}
	\\	&= \frac12 \sum_{\mu=0}^3 \hat\sigma^\mu_j \sum_{\alpha,\beta = 0}^{\chi -1}
    {R_{L\alpha}}^\mu M_{\alpha\beta} \hat y_{R\beta} .
\end{split}\end{align}
\end{subequations}
That is, they depend only on the left four columns and top four rows of $M$, again as desired.

\subsection{Modifying the MPDO}
\label{SS:mod}
Working in the $\{y\}$ bases, we can modify $M_{\alpha\beta}$ for $\alpha, \beta \ge 4$ at will without violating our guarantees.
We wish to do so in a way that reduces the rank of $M$ while doing the least violence, in some sense, to the connected components of correlations across the cut.
Explicitly, we wish to change the quantities $C(A_L,B_R) = \braket{\hat A_L \hat B_R} - \braket{\hat A_L} \braket{\hat B_R}$, where $\hat A_L$, $\hat B_R$ have support on the left, right portions of the chain respectively, as little as possible.

First, let us see what the correlator involves.
Using the definition $\braket{\hat{A}} = \frac1Z \tr \hat{A} \rho$, with $Z \equiv \tr\rho = {R_{L0}}^0 {R_{R0}}^0 M_{00}$, it is easy to see that 
\begin{subequations}
\begin{align}
	\braket{\hat A_L} = \frac{1}{Z} \tr[\hat{y}_{L\alpha} \hat A_L] M_{\alpha\beta} {R_{R\beta}}^0
		= \frac{1}{Z} a_{\alpha} M_{\alpha 0} {R_{R0}}^0
\end{align}
and
\begin{align}
	\braket{\hat B_R} = \frac{1}{Z} {R_{R\alpha}}^0 M_{\alpha\beta} \tr[\hat{y}_{R\beta} \hat B_R]
		= \frac{1}{Z} {R_{L0}}^0 M_{0\beta} b_{\beta} .
\end{align}
\end{subequations}
We define $a_\alpha = \tr[\hat{y}_{L\alpha} \hat A_L]$, $b_\alpha = \tr[\hat{y}_{R\alpha} \hat B_R]$ for convenience.
(Throughout this subsection, we employ Einstein summation notation over the Greek indices.)
The expectation value of the product is then:
\begin{align}\begin{split}
	\braket{\hat A_L \hat B_R} &= \frac{1}{Z} \tr[\hat{y}_{L\alpha} \hat A_L] M_{\alpha\beta} \tr[\hat{y}_{R\beta} \hat B_R]
	\\	&= \frac{1}{Z} a_{\alpha} M_{\alpha\beta} b_{\beta} .
\end{split}\end{align}
Putting all these together we find:
\begin{align}
	C(\hat A_L,\hat B_R) = \frac{1}{Z} a_{\alpha} \left(M_{\alpha\beta} - \frac{M_{\alpha0}M_{0\beta}}{M_{00}}\right) b_{\beta} .
\end{align}
Since we would like to only alter pieces of the density matrix which affect correlations accross the cut at bond $j$, we will modify the matrix in the parenthesis, and denote it by:
\begin{align}
	\tilde{M}_{\alpha\beta} = M_{\alpha\beta} - \frac{M_{\alpha 0}M_{0\beta}}{M_{00}} .
\end{align}

At this point we have pushed one step further the process of re-writing the density matrix so that its structure explicitly reflects the distinction between information we are willing to change and information we are not willing to change, but we still have not truncated it.
To carry out the truncation, perform an SVD on the lower right block of $\tilde M$, writing $\tilde M_{\alpha\beta} = \sum_\gamma X_{\alpha\gamma}r_\gamma Y_{\gamma\beta}$ for $\alpha,\beta \geq 4$.
Choose an integer $\chi'$ (we will see shortly how it relates to the bond dimension of the final truncated MPDO) and insert a projection $P^{(\chi')}$ onto the largest $\chi'$ elements of $r$ to form a new matrix
\begin{align}
  \label{eq:Mreduction}
  \tilde M' = 
  \left[
  \begin{array}{c|ccc}
    \quad&&&\\
    \hline\\
    &&XrP^{(\chi')}Y&\\
    &&&\\
  \end{array}
  \right].
\end{align}
$\tilde M'$ differs from $\tilde M$ only in those elements $\tilde M'_{\alpha\beta}$ with $\alpha, \beta \geq 4$---that is, in those elements that encapsulate correlations with range $\ge 2$.
Moreover, only small elements of $M$ are changed, since we take small (connected) correlations and set them identically to zero.

This truncation results in a new matrix $M'_{\alpha\beta}$ to replace $M_{\alpha\beta}$:
\begin{equation}
	M'_{\alpha\beta} = \tilde{M}'_{\alpha\beta} + \frac{M_{\alpha0} M_{0\beta}}{M_{00}}
\end{equation}
and then 
\begin{equation}
	\rho'= \hat y_{L\alpha} M'_{\alpha\beta} \hat y_{R\beta}.
\end{equation}
The matrix $M'$ has rank at most $8 + \chi'$ (see App.~\ref{app:rank}).%

Since we know $M'$ and (matrix-product representations of) the $x_{L\alpha}$ and $x_{R\beta}$, putting this into MPDO form like \eqref{eq:MulticanonicalMPDO} with bond dimension $8 + \chi'$ is a matter of rearrangement.
To rearrange into MPDO form perform a second singular value decomposition, this time on $M'$, for
\begin{equation}
  M' = U s' V.
\end{equation}
Since $M'$ has rank at most $\chi' + 8$, there will be at most $\chi' + 8$ nonzero singular values $s$.
The density matrix after truncation is then
\begin{equation}
  \rho \mapsto \sum_{\{\mu\}} \big[A_1^{\mu_1} \cdots A_j'^{\mu_j}\big] s' \big[B_{j+1}'^{\mu_{j+1}} \cdots B_{L}^{\mu_L}\big]
		\hat\sigma_1^{\mu_1}\cdots\hat\sigma_L^{\mu_L} ,
\end{equation}
with
\begin{align}\begin{split}
	A'^{\mu_j}_{j} &= A_j Q^*_L U,
\\	B'^{\mu_j}_{j+1} &= V Q^*_L B_{j+1};
\end{split}\end{align}
and the rest of matrices $A_1, \dots, A_{j-1}, B_{j+2}, \dots, B_L$ are untouched.
Note that, regardless of our choice of $\chi'$, the reduced density matrices on sites $1,\dots,j+1$, $j,\dots,L$ are exactly as they were before the truncation.

\section{Results}
We apply the method to a variety of initial states, pure and mixed.
We time evolve by the boustrophedon Trotter decomposition \eqref{eq:trotter} of a Hamiltonian known to satisfy the eigenstate thermalization hypothesis with Trotter step $\delta t = 1.0$ (except where specified).
We work at a maximum bond dimension cutoff (i.e., at each gate application we truncate to this cutoff using the algorithm described in section \ref{S:Method}) and measure performance by varying this cutoff.

\subsection{Hamiltonian}
\label{S:Ham}
We take as our Hamiltonian the spin-1/2 transverse-field Ising model on an $L$-site chain with open boundary conditions:
\begin{equation}
  \label{eq:TFIM}
  H = \sum_{j= 1}^{L-1} S_j^zS_{j+1}^z + \frac12 h^x \sum_{j = 1}^L S^x + \frac12 h^z \sum_{j = 1}^L S^z .
\end{equation}
At $h^z = 0 $ the model is integrable (by Jordan-Wigner and Bogoliubov transformations); the longitudinal field $h^z\sum S^z$ breaks integrability, and at $h^z = 0.8090, h^x = 0.9045$ the model is known to satisfy the eigenstate thermalization hypothesis in a strong sense \cite{PhysRevE.90.052105}.
We work at onsite fields
\begin{equation}
  h^z = 0.8090, \quad h^x = 0.9045
\end{equation}
(except where otherwise specified).
Despite their ETH nature, TFIM Hamiltonians like this can display ill-understood pre-thermalization behavior, thought to be related to long-range emergent conserved or nearly-conserved quantities \cite{PhysRevLett.106.050405, PhysRevE.92.012128, 2016arXiv161004287L}.
We do not expect DMT to be able to capture this emergent-integrable behavior (see Section \ref{S:Int}), so we choose our initial conditions to avoid it.

\subsection{Application: pure-state evolution}
\label{SS:PSE}

\begin{figure}
	\includegraphics[width=0.45\textwidth]{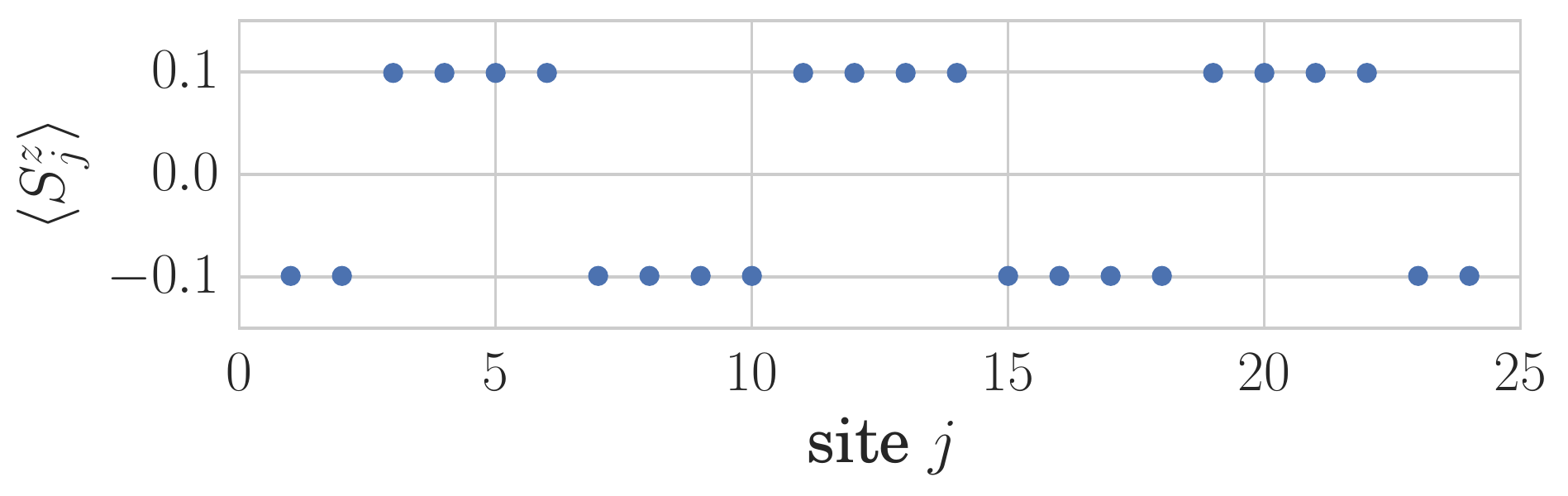}
	\caption{$\langle S^z \rangle$ for the initial state described in \eqref{eq:knearYdef} on a $24$-site chain.}
	\label{fig:PSE:initSz}
\end{figure}

We engineer an initial state with a long-wavelength energy variation by taking a product of $\sigma^y$ eigenstates and rotating blocks of four spins alternately towards $+z$ and $-z$ (cf Fig.~\ref{fig:PSE:initSz}).
The initial state is
\begin{equation}
  \label{eq:knearYdef}
  \knearY = \prod_{j = 1}^L \big[1 + i (1 + g_j) \sigma^+_j\big] \ket{\downarrow\downarrow\dots\downarrow}
\end{equation}
(suitably normalized), where
\begin{equation}
  g_j = 0.1 \times
  \begin{cases}
    -1 & j \bmod 8 = \textrm{1, 2, 7, or 0}, \\
    +1 & j \bmod 8 = \textrm{3, 4, 5, or 6}.
  \end{cases}
\end{equation}
(We choose the state to be near the $\sigma^y$ product state in order that we may avoid the pre-thermalization behavior found in \cite{PhysRevLett.106.050405, PhysRevE.92.012128, 2016arXiv161004287L}.)

Since the initial state is a product state, it may be represented exactly as an MPO with bond dimension $\chi = 1$.
Trotter time evolution increases the bond dimension with each time step $\delta t$, but truncation (whatever the algorithm) kicks in only at a time $t_{\mathrm{trunc}}(\chi_{\mathrm{max}})$ when $\chi(t)$ reaches $\chi_{\mathrm{max}}$.
Thus for each $\chi_{\mathrm{max}}$ the time evolution is semiexact (that is, exact up to error resulting from the Trotter decomposition of the Hamiltonian) for $t < t_{\mathrm{trunc}}(\chi_{\mathrm{max}})$, at which time it begins to deviate from the semiexact value.
This effect appears in all of our results; we also use it to benchmark our method (vide infra).

\begin{figure}
	\includegraphics[width=0.48\textwidth]{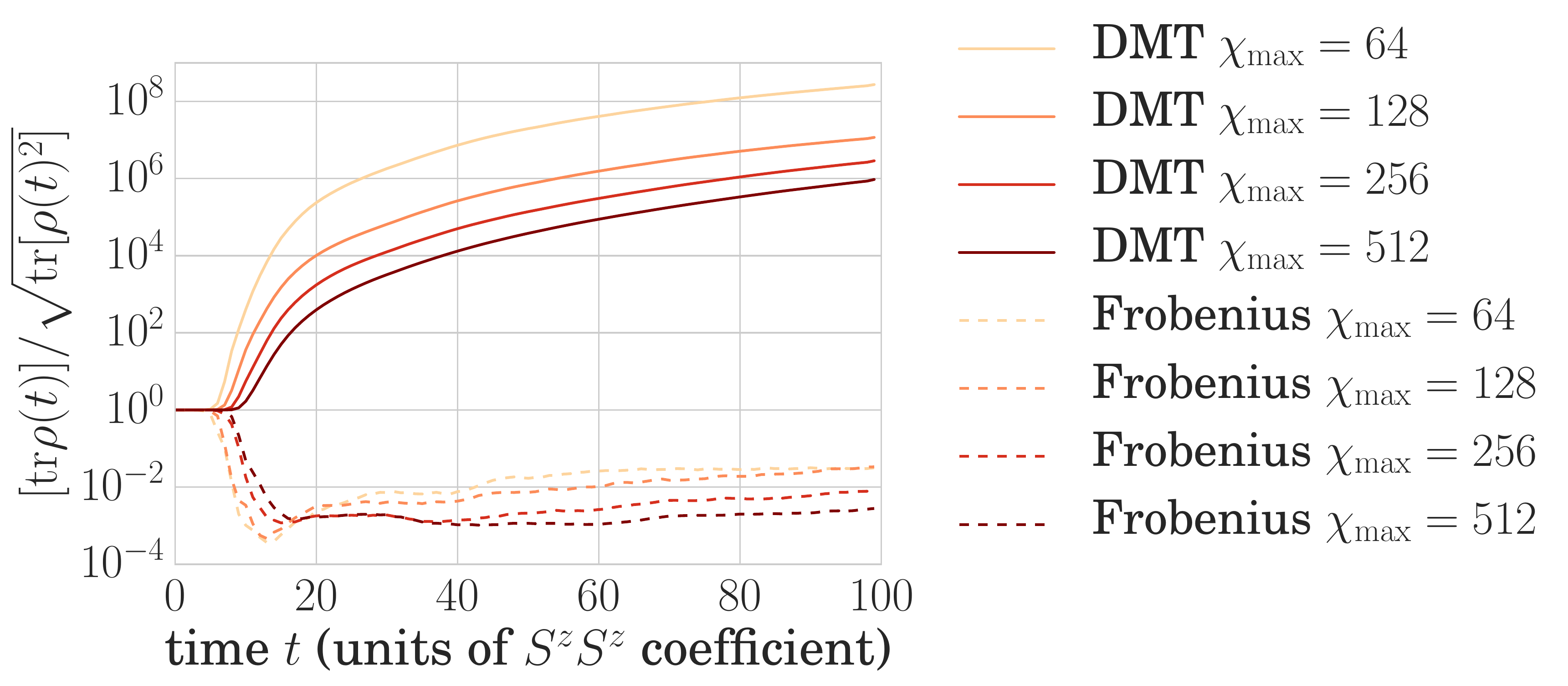}
	\caption{%
		Normalization $Z = [\tr \rho] / \sqrt{\tr [\rho^2]}$ as a function of time, comparing DMT (solid) and Frobenius (dashed), for the initial pure state \eqref{eq:knearYdef} evolving under the Hamiltonian \eqref{eq:TFIM}.
		Note that the second R\'enyi entropy of the whole chain (64-sites long) is $S_2 = 2\ln Z$.
		As $Z = 1$ for a pure state, any deviations from $Z = 1$ result from the truncation.
	}
	\label{fig:PSE:norm}
\end{figure}

Figure~\ref{fig:PSE:norm} shows the normalization $Z = \frac{\tr\rho}{\sqrt{\tr\rho^2}}$ as a function of time.
The normalization is related to the second R\'enyi entropy of the entire chain $S_2 \equiv -\ln \frac{\tr [\rho^2]}{[\tr \rho]^2}$ via $S_2 = 2 \ln Z$.
The DMT produces bath entropy for the system, and this is reflected in the increase of $Z$ as a function of time.
In contrast, we find that the Frobenius method produces non-physical states with $Z < 1$ over the course of time-evolution, which results from negative eigenvalues of the density matrix generated in the truncation.
The observation that $Z \geq 1$ for DMT does not imply positive semi-definitity, but suggests that any error arising from the negative eigenvalues is small and well-controlled.

\begin{figure}
	\includegraphics[width=0.45\textwidth]{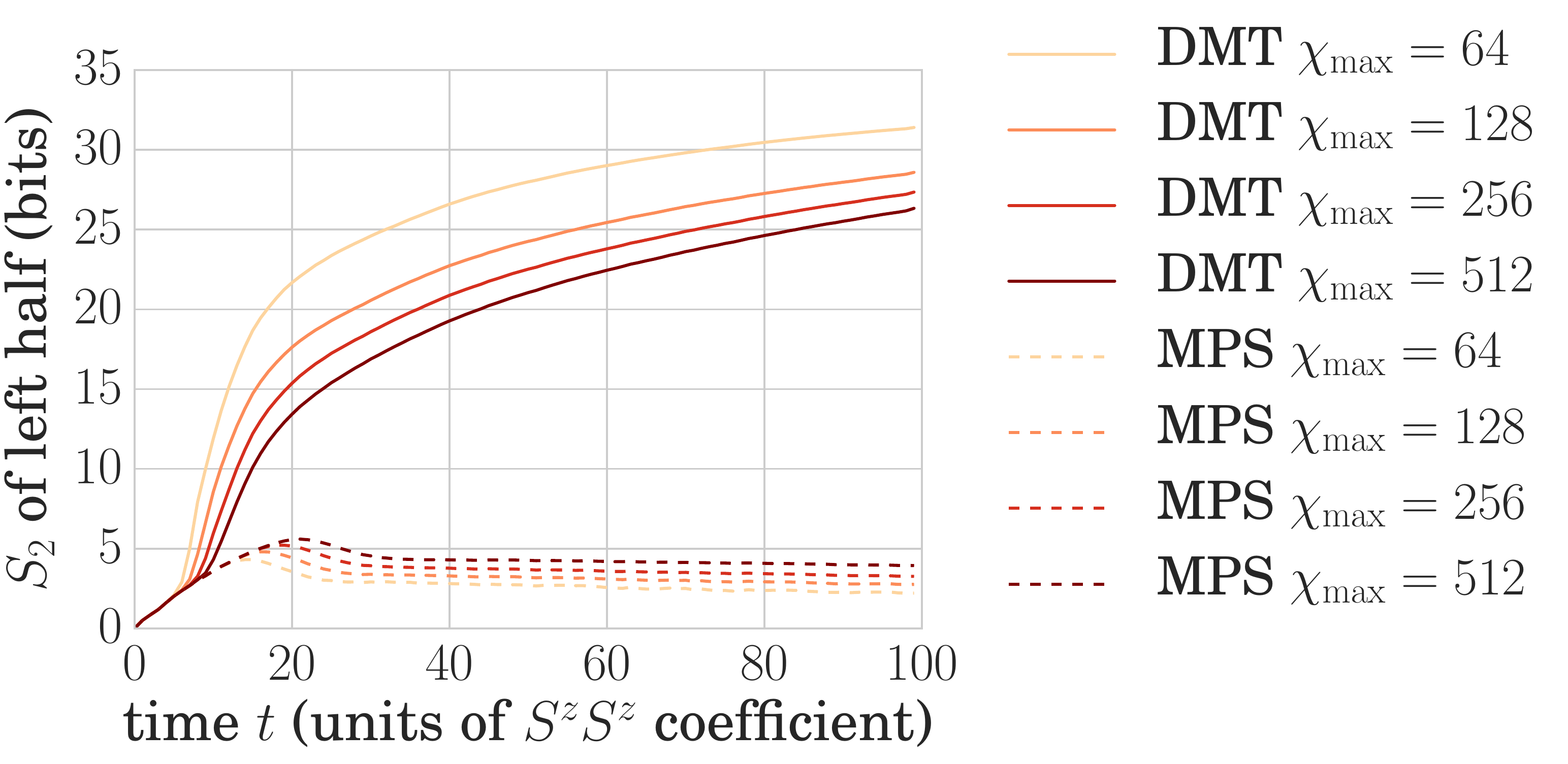}
	\caption{%
	Second R\'enyi entropy (in units of bits) of left half of a 64-site chain for an initial pure state \eqref{eq:knearYdef} evolving under the Hamiltonian \eqref{eq:TFIM}, in MPS and DMT simulations.
	The largest entropy we see ($\chi = 64$ at $t=100$) is $S_2 \approx \unit[31.4]{bits}$, very close to the theoretical maximum of \unit[32]{bits}.
	}
	\label{fig:PSE:Renyi2}
\end{figure}

Figure~\ref{fig:PSE:Renyi2} shows the second R\'enyi entropy of the left half of the chain---that is, the subsystem consisting of sites $1$ to $L/2=32$.
(The von Neumann entropy is difficult to calculate for MPDOs, while second R\'enyi entropies are nearly trivial.)
In contrast to MPSs, MPDOs can represent states with arbitrarily large entropy by replacing system entanglement entropy with bath entropy. 

Note that once truncation starts, the entropy in the DMT simulation increases above that in MPS.
This is not unexpected: in ordinary MPS TEBD the entanglement entropy of the left half is exactly its entropy with the right half and is a property of the MPS at bond $\lfloor L/2 \rfloor$, so it can only increase when we apply a gate at bond $\lfloor L/2 \rfloor$.
In the DMT algorithm, on the other hand, the entanglement entropy of the left half of the chain is entanglement entropy not only with the right half but also with a notional bath, and it increases with every truncation on bonds within the left half.

\begin{figure}
	\includegraphics[width=0.41\textwidth]{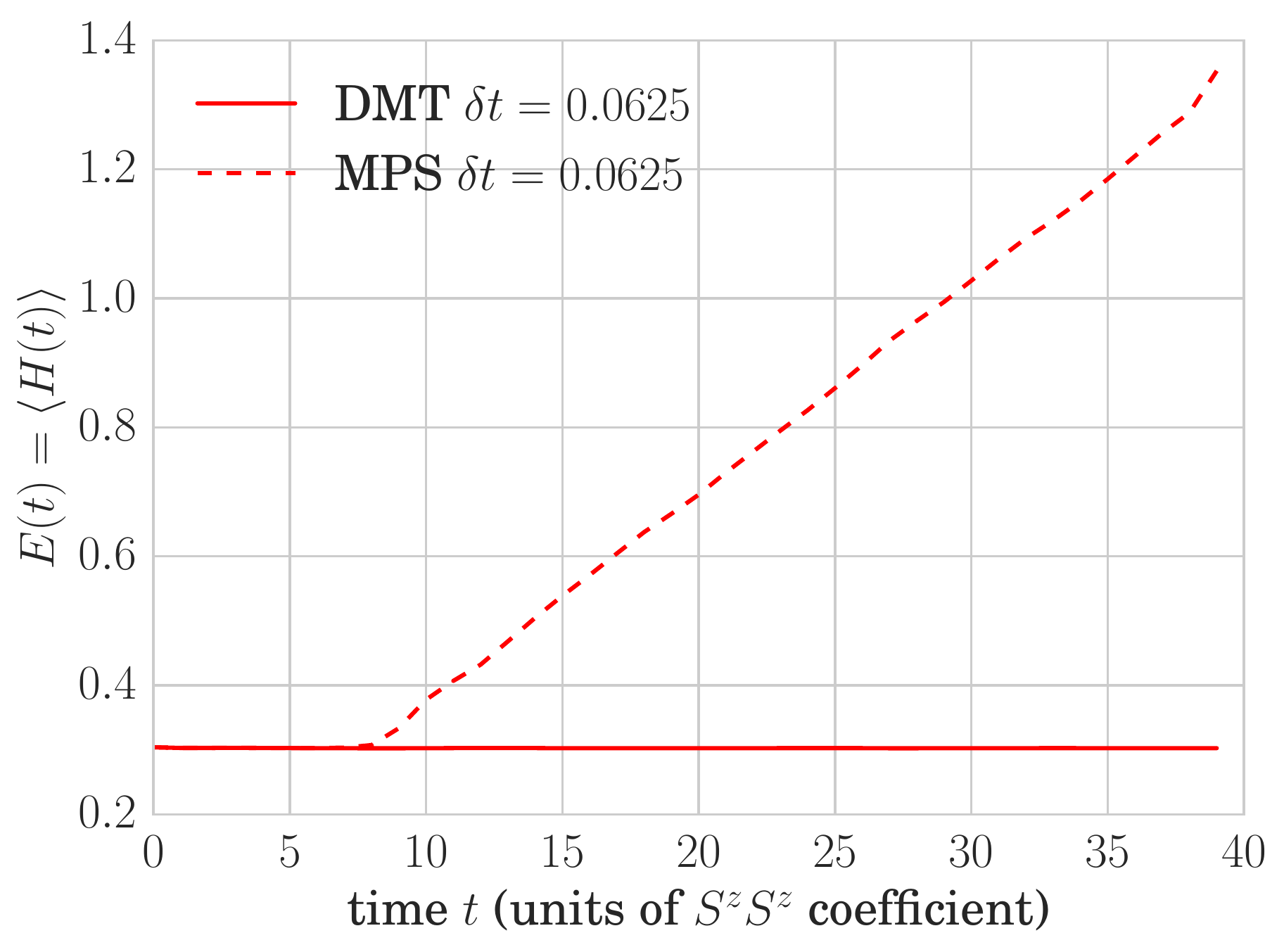}
	\caption{Energy over time at fixed $\chi = 16$ for the initial pure state \eqref{eq:knearYdef} evolving under the Hamiltonian \eqref{eq:TFIM} at system size $L = 64$.
		By design, the total energy remain constant under DMT.}
	\label{fig:PSE:E}
\end{figure}

Figure \ref{fig:PSE:E} shows the system's total energy over time as simulated by ordinary TEBD and our density-matrix TEBD. 
In the DMT simulation, the energy is constant.
MPS time evolution, however, imposes an additional `heating' whereas DMT is designed to conserve total energy.
Because the MPS representation biases towards low-entanglement states, the system drifts towards the extrema of the energy spectrum over time.
(The simulation begins with positive energy, and hence drifts towards the negative temperature state $T \rightarrow 0^-$.)

Onsite spins are easy to measure but hard to analyze: their expection values are noisy and they decay quickly with time.
Instead, we measure a Fourier component of the energy density.
The energy density $\epsilon_j$, defined over a pair of sites, is
\begin{align}
  \label{eq:epsdef}
  \epsilon_1 &= \frac{h^z}{2} \big(S^z_1 + \tfrac12 S^z_2\big) + \frac{h^x}{2} \big(S^x_1 + \tfrac12 S^x_2\big) + S^z_1 S^z_2 ,  \notag\\
  \epsilon_{ 1 < j < L-1 } &=   \frac{h^z}{2} \big(\tfrac12 S^z_j + \tfrac12 S^z_{j+1}\big) + (x \leftrightarrow z) + S^z_jS^z_{j+1} ,  \notag\\
  \epsilon_{L-1} &= \frac{h^z}{2}\big(\tfrac12 S^z_{L-1} + S^z_{L}\big) + (x \leftrightarrow z) + S^z_{L-1}S^z_{L} .
\end{align}
We measure a Fourier component of the energy density
\begin{equation}
  \label{eq:EpsFourierDef}
  \epsilon_{k = \pi/4} = - \frac 1 L \sum_{j = 1}^{L-1}e^{ikj} \epsilon_j, \qquad k = \pi/4
\end{equation}
with a wavelength of 8 sites.
Fourier components are eigenmodes of the diffusion equation which should govern the system's long-time non-equilibrium behavior.
We choose this particular component (and choose the initial state accordingly) because its wavelength is long enough that it should not immediately decay, but not so long as to be longer than accessible system sizes.

\begin{figure*}
	\begin{minipage}{\textwidth}
		\includegraphics[width=0.99\textwidth]{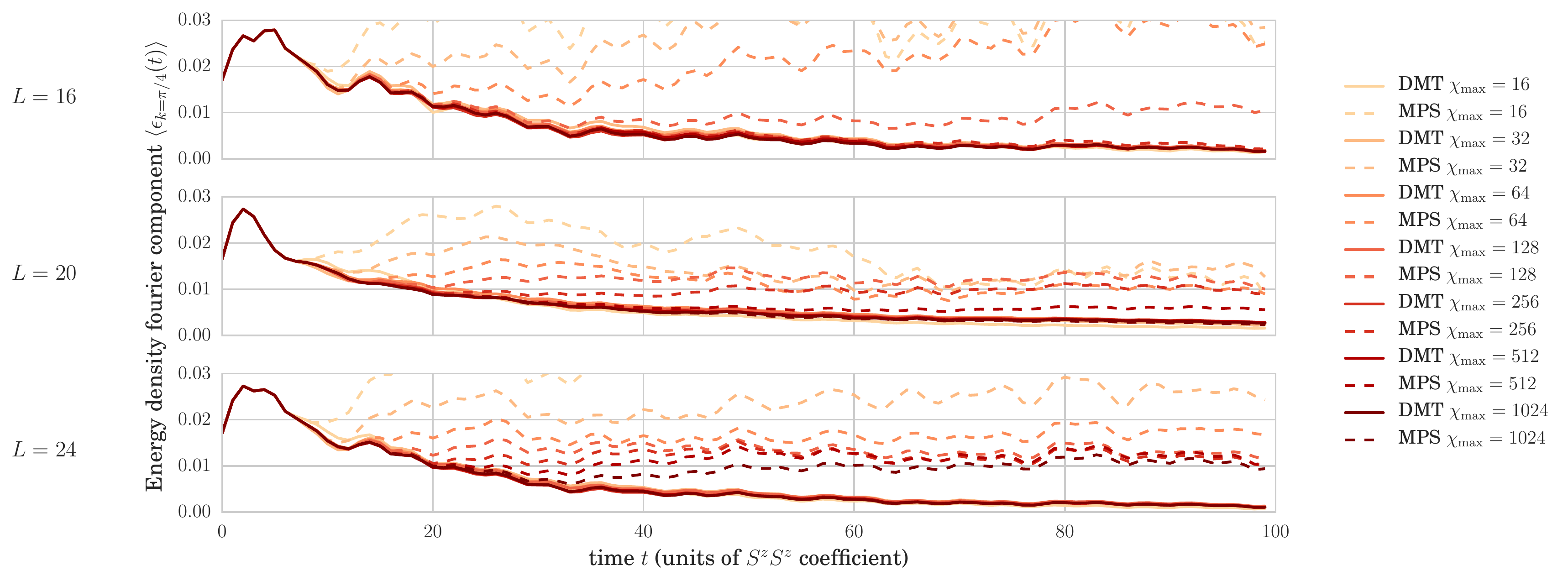}
		\\[-2ex]\hspace{-\textwidth}\begin{minipage}{0mm}\vspace{-125mm}\subfigure[]{\label{fig:PSE:epsFC:abs}}\hspace{-27mm}\end{minipage}
	\end{minipage}
	\begin{minipage}{\textwidth}
		\includegraphics[width=0.99\textwidth]{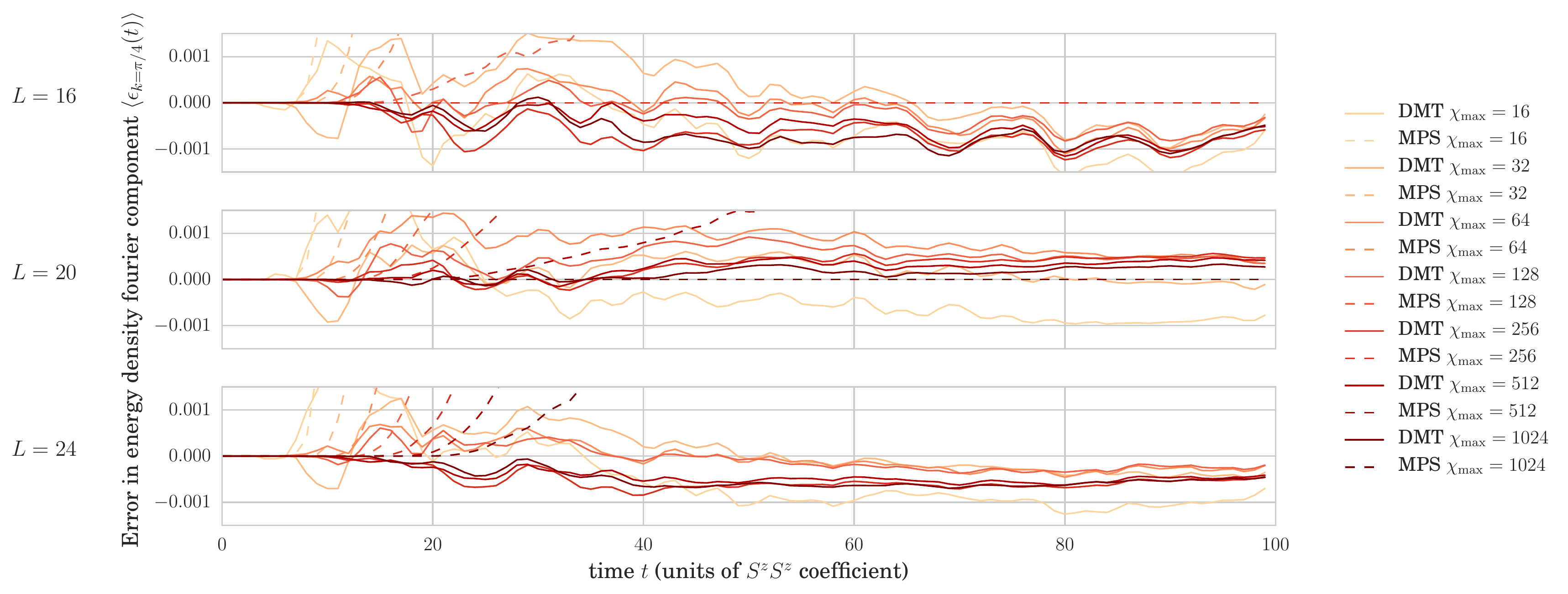}
		\\[-2ex]\hspace{-\textwidth}\begin{minipage}{0mm}\vspace{-125mm}\subfigure[]{\label{fig:PSE:epsFC:err}}\hspace{-27mm}\end{minipage}
	\end{minipage}
	\caption{%
		(a) Fourier component of energy density for the initial pure state \eqref{eq:knearYdef} evolving under the Hamiltonian \eqref{eq:TFIM} on chains of length $L = 16, 20, 24$. 
		(b) The `error' of the energy density,
		measured by comparing each data set with the semiexact result as simulated by MPS at $\chi = 2^{L/2}$.
		Note we do not show the MPS simulations for $\chi_{\mathrm{max}} = 2048, 4096$.
	}
	\label{fig:PSE:epsFC}
\end{figure*}

Figure~\ref{fig:PSE:epsFC:abs} shows the the Fourier component of energy density \eqref{eq:EpsFourierDef} in as simulated by MPS (dashed) and DMT for $L = 16$, $20$, $24$.
At fixed maximum bond dimension $\chi \ll 2^{\lfloor L/2\lfloor}$, DMT is more accurate than MPS TEBD, which illustrates the power of DMT in both short- and long-time dynamics.
Moreover, where MPS TEBD error increases with system size, DMT error decreases.
This is due to finite-size deviations from thermalizing behavior: oscillations about the local equilibrium values for local operators result from long-range coherences that we do not expect to be able to capture.

Any pure state on a system of length $L$ can be represented exactly by matrix product states with bond dimension $\chi_{\mathrm{max}} = 2^{\lfloor L/2 \rfloor}$.
At this bond dimension the evolution by MPS TEBD becomes semiexact for all times; no truncation occurs for the simulation.
We can therefore simulate pure-state evolution of an MPS using exactly the same Hamiltonian~\eqref{eq:TFIM} and boustrophedon Trotter decomposition~\eqref{eq:trotter}, and compare the results to those of DMT, shown in Fig.~\ref{fig:PSE:epsFC:err}.
The data is a measure of the error introduced by the truncation; these are small $\approx 10^{-3}$ for a wide range of bond dimensions in the DMT simulations.

\subsection{Application: mixed-state evolution (near equilibrium)}
\label{SS:MSnEE}
To probe the behavior of our algorithm near equilibrium, we take as our initial state a Gibbs state with a spatially-varying temperature
\begin{equation}
  \label{eq:MSnE}
  \rho_0 \propto \exp\bigg[\sum_j \beta_j \epsilon_j\bigg]
\end{equation}
with $\epsilon_j$ the energy density of Eq.~\eqref{eq:epsdef}, and
\begin{equation}
  \beta_j = \beta_0 \big(1 + g'_j\big)
\end{equation}
where
\begin{equation}
  g'_j = 0.1 \times
  \begin{cases}
    0 & j \bmod 8 = \textrm{1, 2, 7, or 0}, \\
    1 & j \bmod 8 = \textrm{3, 4, 5, or 6}.
  \end{cases}
\end{equation}
This temperature profile is broadly similar to the $S^z$ profile we impose on the pure initial state (see Eq.~\eqref{eq:knearYdef} and Fig.~\ref{fig:PSE:initSz}).

See Appendix \ref{APP:GIBBS} for details of the construction of the Gibbs state.

\begin{figure}
	\begin{minipage}{83mm}
 		\includegraphics[width=80mm]{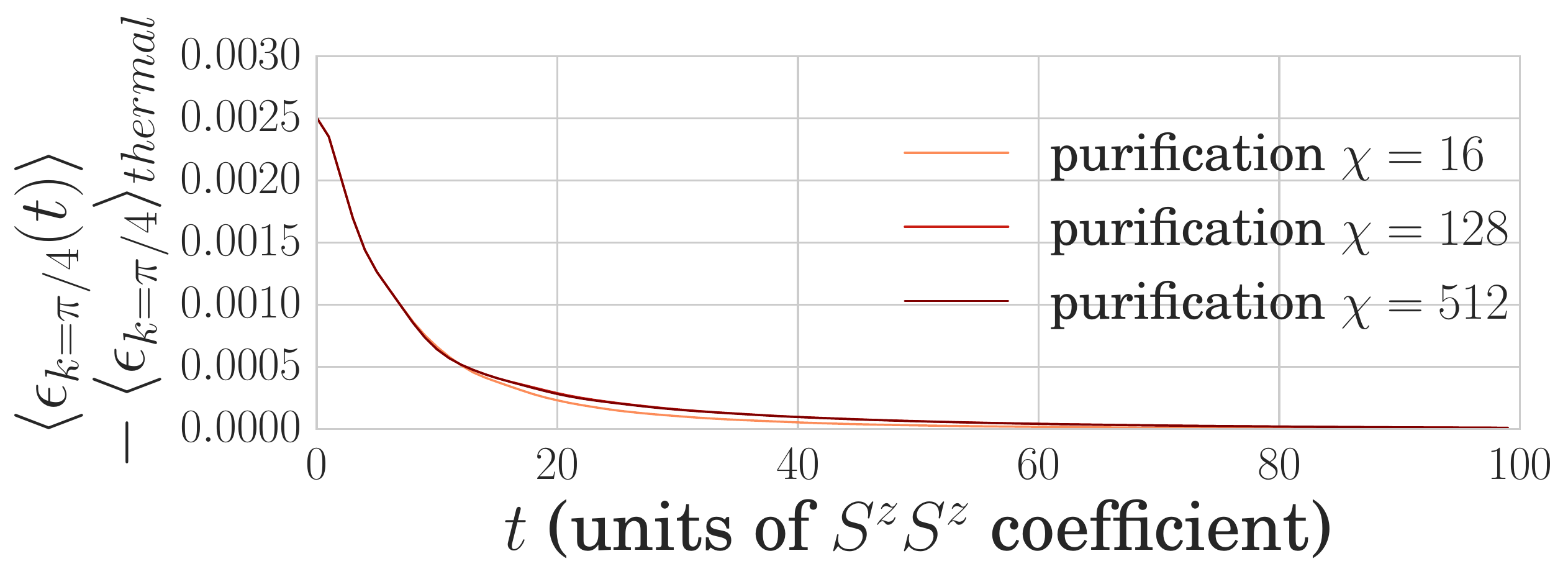}
		\\[-2ex]\hspace{-\textwidth}\begin{minipage}{0mm}\vspace{-61mm}\subfigure[]{\label{fig:MSnEE:epsfc:KBM}}\hspace{-27mm}\end{minipage}
	\end{minipage}
	\begin{minipage}{83mm}
		\includegraphics[width=80mm]{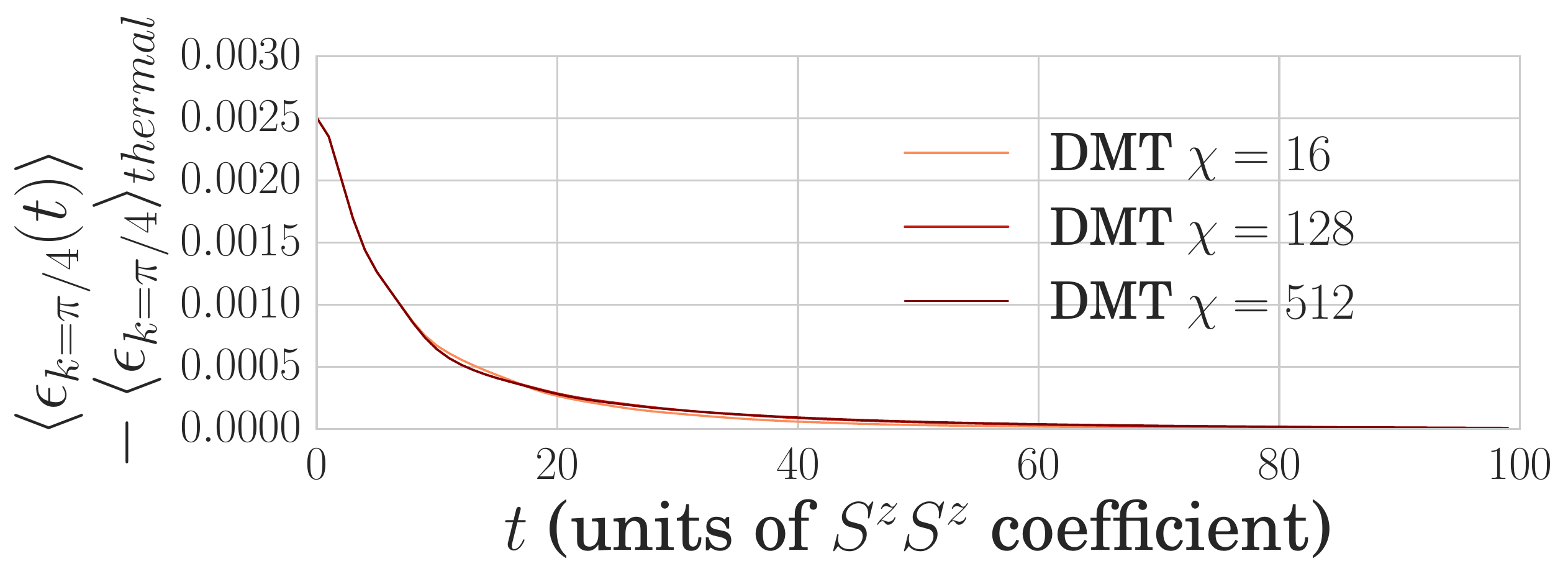}
		\\[-2ex]\hspace{-\textwidth}\begin{minipage}{0mm}\vspace{-61mm}\subfigure[]{\label{fig:MSnEE:epsfc:DMT}}\hspace{-27mm}\end{minipage}
	\end{minipage}
	\caption{Fourier component of energy density for (a) purification time evolution and (b) DMT for the near-equilibrium mixed state \eqref{eq:MSnE} evolving under the Hamiltonian \eqref{eq:TFIM} on a 128-site chain.
		The thermal value is $\langle \epsilon_{k = \pi/4} \rangle_{\mathrm{thermal}} = -0.00038$}
	\label{fig:MSnEE:epsfc}
\end{figure}

\begin{figure}
	\begin{minipage}{83mm}
		\includegraphics[width=80mm]{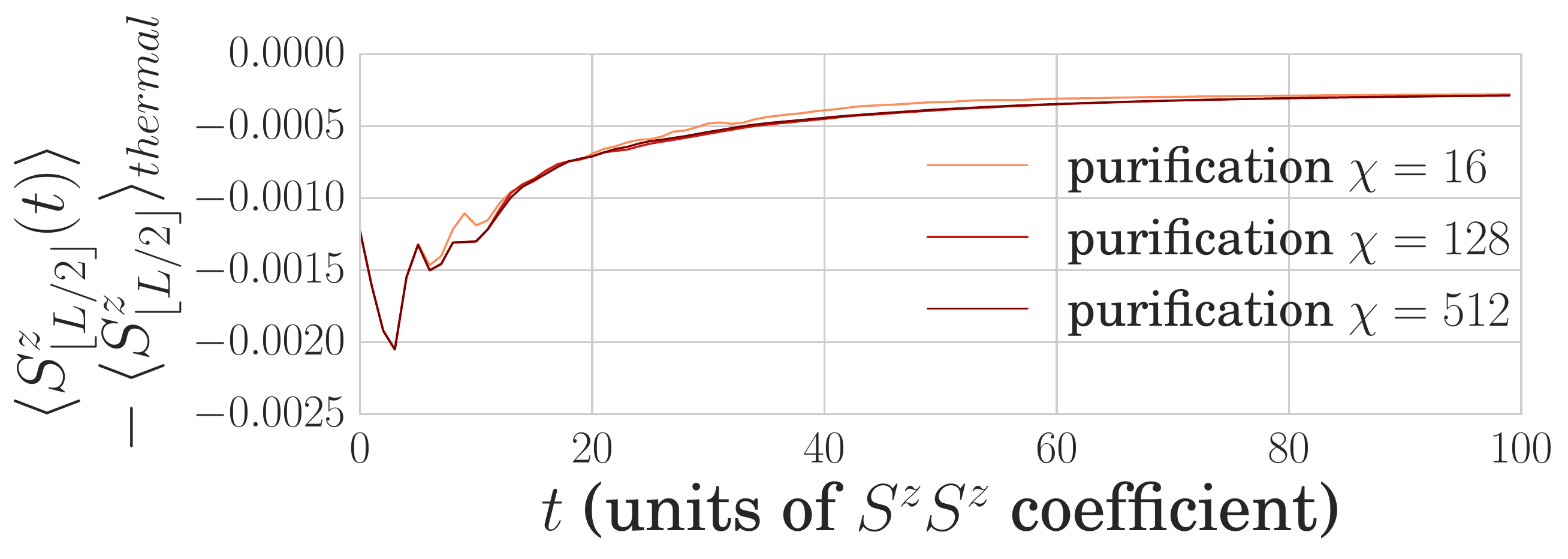}
		\\[-2ex]\hspace{-\textwidth}\begin{minipage}{0mm}\vspace{-61mm}\subfigure[]{\label{fig:MSnEE:Sz:KBM}}\hspace{-27mm}\end{minipage}
	\end{minipage}
	\begin{minipage}{83mm}
		\includegraphics[width=80mm]{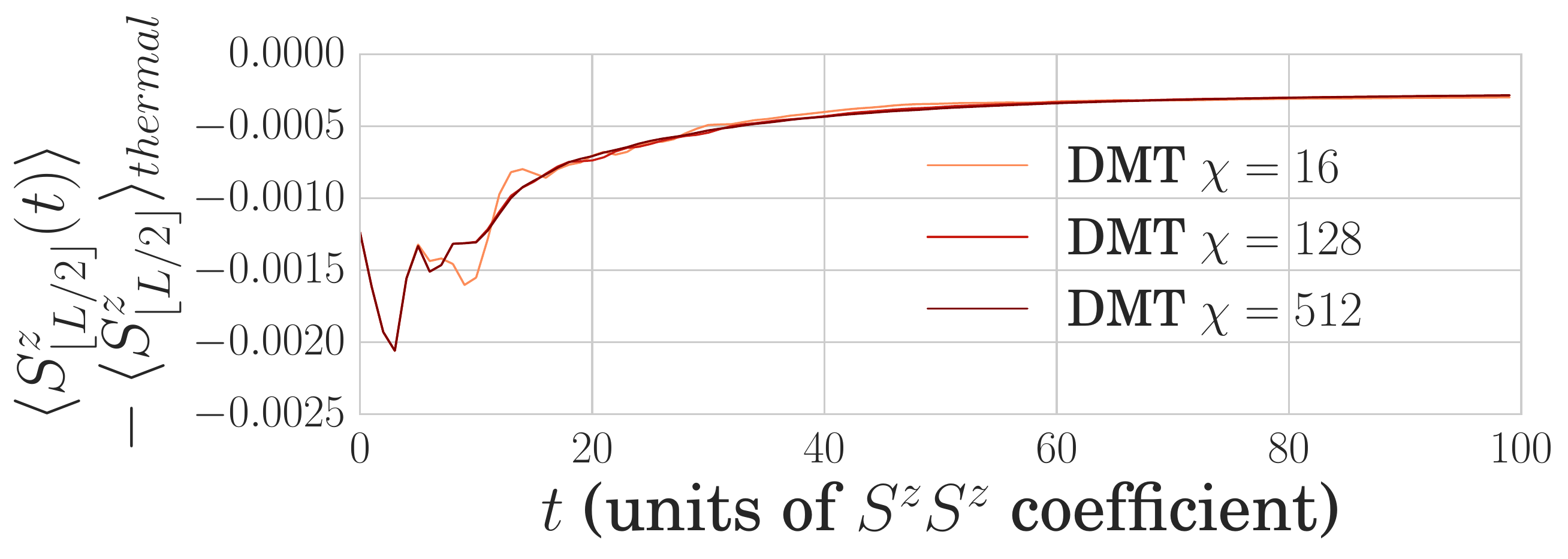}
		\\[-2ex]\hspace{-\textwidth}\begin{minipage}{0mm}\vspace{-61mm}\subfigure[]{\label{fig:MSnEE:Sz:DMT}}\hspace{-27mm}\end{minipage}
	\end{minipage}
	\caption{Expectation value of $S^z$ at the midpoint of the chain for (a) purification time evolution and (b) DMT for the near-equilibrium initial state \eqref{eq:MSnE} evolving under the Hamiltonian \eqref{eq:TFIM} on a 128-site chain.
		Both methods converge very quickly, so they give nearly identical results (cf.\ Fig.~\ref{fig:APP:MSnEE:conv:Sz}). 
		This expectation value fails to approach the thermal value due to the large Trotter step we use ($dt = 1.0$).
		The thermal value is $\langle S^z_{\lfloor L/2 \rfloor} \rangle_{\mathrm{thermal}} = -0.0622$.
	}
	\label{fig:MSnEE:Sz}
\end{figure}

In Figs.~\ref{fig:MSnEE:epsfc} and \ref{fig:MSnEE:Sz} we compare DMT to the purification method of Karrasch, Bardarson and Moore \cite{KBM2013}, which we label ``purification''.
This method takes advantage of the freedom to apply unitaries to the ancillae by time-evolving the ancillae backwards even as it evolves the physical system forwards.
The time-evolution framework is therefore very similar to ours; the chief differences are in the interpretation of the vector space $(\mathbf{C}^{d^2})^L$ in which one works and in the truncation algorithm.
The similarity is magnified by our choice to use the boustrophedon Trotter decomposition \eqref{eq:trotter} not only for DMT but for purification. 

Both DMT and this purification time evolution converge very quickly, as one might expect: the results are essentially identical between the methods and between bond-dimension cutoffs, even to quite small bond dimensions.
(Note that we subtract the thermal value in each case.)

\section{Conclusion}
We have presented an algorithm for approximating density operators by low-rank matrix product operators suitable for simulating long-time quantum dynamics.
The method exactly preserves expectation values of operators on up to three contiguous sites,
and it slots neatly into a standard Trotter-decomposition framework for time evolution of matrix product structures (TEBD),
allowing time evolution by an ETH Hamiltonian of a variety of initial states.

Our algorithm, DMT, qualitatively outperforms its nearest competitor (ordinary MPS TEBD) for pure initial states.
We use the fact that matrix product density operators with small bond dimension can represent states with high entropy to circumvent the area-law entanglement bound on matrix product states.
Thus far the work is unoriginal: Zwolak and Vidal realized this was possible more than a decade ago.
Our key insight is that we can preserve the trace of the density matrix and the expectation values of conserved quantities by appropriately rotating the Schmidt spaces at the bond at which we truncate.
Consequently, DMT can simulate time evolution by ETH Hamiltonians to arbitrary times using memory and computation time polynomial in system size.

In addition, DMT matches the current state of the art (purification time evolution) for near-equilibrium mixed initial states and outperforms it for far-from-equilibrium initial states.

We did not compare our algorithm to the interesting recent work of Leviatan et al.\cite{2017arXiv170208894L}.
They use intuition not unlike ours to argue that the time-dependent variational principle of Haegeman et al. \cite{haegeman_time-dependent_2011, haegeman_unifying_2016},
which approximates quantum dynamics as a classical Hamiltonian flow on the manifold of low-bond-dimension matrix product states and therefore exactly conserves energy,
should give good results for ETH systems.
The goals of Ref.~\onlinecite{2017arXiv170208894L}---and, consequently, their benchmarking protocols---differ from ours, but
a direct comparison could be an interesting topic for future work.

The reader would be right to worry that our method does not converge: as we increase the bond dimension above a certain value (perhaps $2^5\mbox{--}2^6$), the accuracy of our method does not improve.
We suspect that---once again---this is a result of the operators $O^y_{l,t}$, whose large expectation values result from the fact that we start near an $S^y_l$ eigenstate.
When we reduce the rank of the matrix $\tilde M$ in \eqref{eq:Mreduction}, we still do so in a way that minimizes error with respect to the Frobenius norm (even though we have arranged to exactly preserve very-short-range operators).
This means that the operators $O^y_{l,t}$ again dominate the error, and the matrix resulting from the truncation is pulled toward those operators.
The obvious next step is to reduce rank in such a way that we minimize error with respect to a different norm, one that takes into account the spatial structure of the operator space: if we truncate at bond $j$, we should weight errors along $\sigma^z_{j - 1}\sigma^z_{j+2}$ more heavily than errors along $\sigma^z_{j - 7}\sigma^z_{j+6}$.
Such controlled-metric truncation is a natural extension of this work.

One natural question to attack using our algorithm is the characterization of the ergodic side of the MBL transition. The random field Heisenberg model with small disorder appears to satisfy the ETH \cite{PhysRevB.82.174411}, but the nature of its dynamics is unclear (see the review of Luitz and Bar Lev \cite{2016arXiv161008993L}).
Quantities like the spin-spin correlation $\langle S^z_{i+r}(t) S^z_i(t) \rangle$, from which one can compute a number of diagnostics for subdiffusion, should be straightforward to calculate using our method.

More interesting still are questions about interfaces between ETH and MBL systems.
Besides being of inherent interest (how large must a bath be to thermalize an MBL system of a given size? 
How quickly does it thermalize?), answers to these questions will shed light on the phenomenological RG schemes of Potter, Vasseur, and Parameswaran \cite{PhysRevX.5.031033} and Vosk, Huse, and Altman \cite{PhysRevX.5.031032} for which ETH-MBL interfaces are fundamental building blocks.
Because MBL systems display low entanglement in a wide variety of situations, we expect our algorithm to be able to simulate both bath and system out to large system sizes.

\acknowledgments
We acknowledge support of the Caltech Institute for Quantum Information and Matter, an NSF Physics Frontiers Center supported by the Gordon and Betty Moore Foundation.
CDW acknowledges the generous support of the National Science Foundation Graduate Research Fellowship under Grant No. DGE‐1144469.
GR acknowledges the generous support of the Packard Foundation, and the NSF through award DMR-1410435.

\clearpage
\appendix

\section{Time-evolution framework}
\label{APP:TEF}
Consider a Hamiltonian
\begin{equation}
  H = \sum H_j
\end{equation}
(for instance \eqref{eq:TFIM}) where $H_j$ is supported on sites $j, j+1$.
Call the onsite Hilbert space $\mathcal H_j$ and its dual $\mathcal H_j^*$.
Pure states then live in a Hilbert space $\mathcal{H} = \mathcal H_j^{\otimes L}$, and density operators $\mathcal{\rho}$ in a Hilbert space $\mathcal H \otimes \mathcal H^* = [\mathcal H_j \otimes \mathcal H_j^*]^{\otimes L}$.
Closed-system Hamiltonian evolution, then, is
\begin{equation}
  \frac{d}{dt} \rho = -i[H,\rho] \equiv -i H^\sharp \rho
\end{equation}
with a linear superoperator Hamiltonian defined
\begin{align}\begin{split}
	H^\sharp &: \mathcal H \otimes \mathcal H^* \to \mathcal H \otimes \mathcal H^*,
\\	H^\sharp &= H \otimes 1 - 1 \otimes H.
\end{split}\end{align}
For a spin-1/2 chain, we can write $S^{x,y,z}_j$ for operators on the ordinary (``ket'') space $\mathcal H_j$ and $T^{x,y,z}_j$ for operators on the dual (``bra'') space $\mathcal H_j^*$; in this notation, the superoperator corresonding to our fruit-fly Ising Hamiltonian \eqref{eq:TFIM} is 
\begin{align}
  H^\sharp  &= \sum H^\sharp_j\notag\\
  &\equiv \sum \Big[(S^z_jS^z_{j+1} - T^z_jT^z_{j+1})\notag \\
  & \qquad\quad + \frac12 h^x (S^x_j - T^x_j) + \frac12 h^z (S^z_j - T^z_j)\Big] .
\end{align}

In order to time-evolve a density matrix by a time $t$, one applies the superoperator unitary $e^{-iH^\sharp t }$. We discretize this operator by a timestep $\delta t$, as usual:
\begin{equation}
  e^{-iH^\sharp t} = \Big[e^{-iH^\sharp\delta t}\Big]^{t/\delta t} .
\end{equation}
We then perform a second-order Trotter decomposition into $2L - 1$ two-site unitaries (``gates'')
\begin{equation}
  \label{eq:trotter}
  e^{-iH^\sharp\delta t}
  \simeq
  \underbrace{\prod_{j = 1}^{L-1}\Big[e^{-iH^\sharp_j\delta t/2}\Big]}_{\text{leftward sweep after}}
  \underbrace{\prod_{j = L-1}^{1}\Big[e^{-iH^\sharp_j\delta t/2}\Big]}_{\text{rightward sweep}};
\end{equation}
the error is of order $L\|h_j\|^3 \delta t^3$, where $\|h_j\|$ is an estimate of the typical magnitude of the terms $h_j$.
(Note that this is a boustrophedon, DMRG-like ``sweep'', not the usual even-odd Trotter decomposition of Vidal's TEBD.
We choose this Trotter decomposition for reasons of numerical stability.)

We apply this series of gates to a so-called matrix product density operators (MPDO): a representation of a density matrix $\rho$ of the form
\begin{equation}
  \label{eq:MulticanonicalMPDO}
  \rho = \sum \big[A_1^{\mu_1}\cdots A_j^{\mu_{j-1}}\big] s \big[B_{j}^{\mu_{j}} \cdots B_{L}^{\mu_L}\big]\hat \sigma_1^{\mu_1}\cdots\hat\sigma_L^{\mu_L}.
\end{equation}
The common dimension of the matrices $A_{n}^{\mu_n}, A_{n+1}^{\mu_n}$ (for any $n$) is called the \textit{bond dimension}; the bond dimension controls the time and memory requirements for storing and operating on the MPDO.
(This MPDO is in multi-canonical form with orthogonality center at site $j$.
All MPDOs with which we work will be in multicanonical form.)
It is convenient to think of applying a gate (say, $e^{-ih^\sharp_{j, j+1}\delta t/2}$) as moving the orthogonality center from a bond adjacent to the bond on which the gate is applied (in this case bond ${j+1}$) to the bond on which it was applied; the boustrophedon Trotter decomposition \eqref{eq:trotter} thus drags the orthogonality center from right end to left and back, over and over again.

When applied to an MPDO, each two-site gate increases the bond dimension at bond by up to a factor of $(\dim \mathcal H)^2$.
Consequently, we must \textit{truncate}: approximate the MPDO by another with a smaller bond dimension (and hence less demanding memory and time requirements);
section \ref{S:Method} describes our truncation algorithm.

\section{Relation between truncation of correlation matrix and bond dimension of MPDO}
\label{app:rank}
In Sec.~\ref{SS:mod}, we truncated
\begin{equation}
  \tilde M \mapsto \tilde M'
\end{equation}
such that the block $\tilde M'_{\alpha\beta}, \alpha,\beta \ge 4$ has rank $\chi'$, then claimed that
\begin{equation}
  \rank \Bigg[ M'_{\alpha\beta} = \tilde M'_{\alpha\beta} + \frac{M_{\alpha 0}M_{0\beta}}{M_{00}}\Bigg] \le \chi' + 8\;.
\end{equation}

To see that this is true, first decompose $\tilde M'$ as
\begin{equation}
  \tilde M' = \tilde M'^A + \tilde M'^B + \tilde M'^C
\end{equation}
with $\tilde M'^{A,B,C}$ left, upper, and lower right blocks of $M'$ respectively:
\begin{align}
  \begin{split}
  \tilde M'^A_{\alpha\beta} &= \tilde M'_{\alpha\beta}, 0 \le \alpha \le 3\\
  \tilde M'^B_{\alpha\beta} &= \tilde M'_{\alpha\beta}, 3 < \alpha, 0 \le \beta \le 3\\
  \tilde M'^C_{\alpha\beta} &= \tilde M'_{\alpha\beta}, 3 < \alpha, \beta
  \end{split}
\end{align}
(other elements zero). These have ranks
\begin{align}
  \begin{split}
  \rank\tilde{M}'^A &\le 4 ,  \\
  \rank\tilde{M}'^B &\le 4 ,  \\
  \rank\tilde{M}'^C &\le \chi' ,
  \end{split}
\end{align}
so
\begin{align}\begin{split}
	\rank(\tilde M') &\le \rank\tilde{M}'^A + \rank\tilde{M}'^B + \rank\tilde{M}'^C \\&\le 8 + \chi'\;.
\end{split}\end{align}
Since
\begin{equation}
  \range \frac{M_{\alpha 0}M_{0\beta}}{M_{00}} \subseteq \range \tilde M'^A
\end{equation}
(the range of an operator is also known as its column space)
we have
\begin{equation}
  \rank M' = \rank \tilde M' \le \chi' + 8,
\end{equation}
as desired.

\section{Matrix product density operator representations of Gibbs states}
\subsection{Construction}
\label{APP:GIBBS}
In sections \ref{SS:MSnEE} and \ref{SS:MSffEE} we require a Gibbs MPDO as our initial state.
The Gibbs state is
\begin{equation}
  \rho \propto e^{-\beta H} = e^{-\beta H/2} I e^{-\beta H/2};
\end{equation}
which is precisely the imaginary-time evolution of the product MPDO $I$ by the Hamiltonian superoperator
\begin{equation}
  H_{\mathrm{therm}}^\sharp = H \otimes 1 + 1 \otimes H
\end{equation}
out to time $\beta/2$.
We approximate this imaginary-time evolution not by tDMRG with the boustrophedon Trotter decomposition \eqref{eq:trotter}, but by ordinary TEBD using the trick of Hastings for numerical stability (as described in 7.3.2 of Schollw\"ock's magisterial review\cite{Schollwock201196}, q.v.).

\subsection{Estimating thermal expectation values}
In analyzing the time evolution of ETH states, one naturally requires Gibbs state expectation values as a function of total energy (or, equivalently, energy density): the long time limit of an expectation value is given by its expectation value in a Gibbs state whose energy density matches that of the initial state.
We tabulate energy densities and observables of interests for Gibbs states at a variety of temperatures using MPDOs as described above;
To find the long-time limit of an expectation value, we measure the energy density of the initial state and linearly interpolate between the two nearest Gibbs energy densities.
Note that this does not account for Trotter heating (that is, the fact that---because the system actually simulated is a Floquet system with period given by the Trotter step $\delta t$, its energy as measured by the Hamiltonian simpliciter gradually increases).

\section{Application: mixed-state evolution (far from equilibrium)}
\label{SS:MSffEE}
One might worry that the two initial states \eqref{eq:knearYdef} from Sec. \ref{SS:PSE} and \eqref{eq:MSnE} from Sec. \ref{SS:MSnEE} are each special cases in their own ways: the first is a pure state, and the second is very near equilibrium.

\begin{figure}
	\begin{minipage}{83mm}
		\includegraphics[width=80mm]{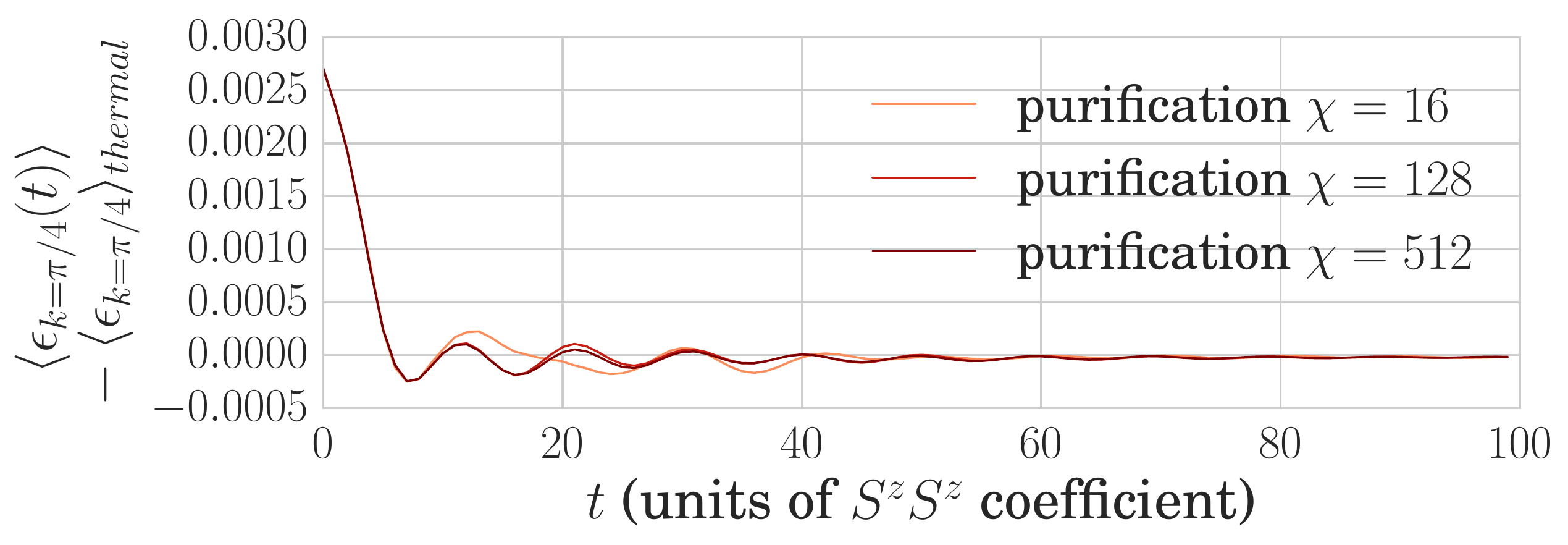}
		\\[-2ex]\hspace{-\textwidth}\begin{minipage}{0mm}\vspace{-52mm}\subfigure[]{\label{fig:MSffEE:epsfc:KBM}}\hspace{-27mm}\end{minipage}
	\end{minipage}
	\begin{minipage}{83mm}
		\includegraphics[width=80mm]{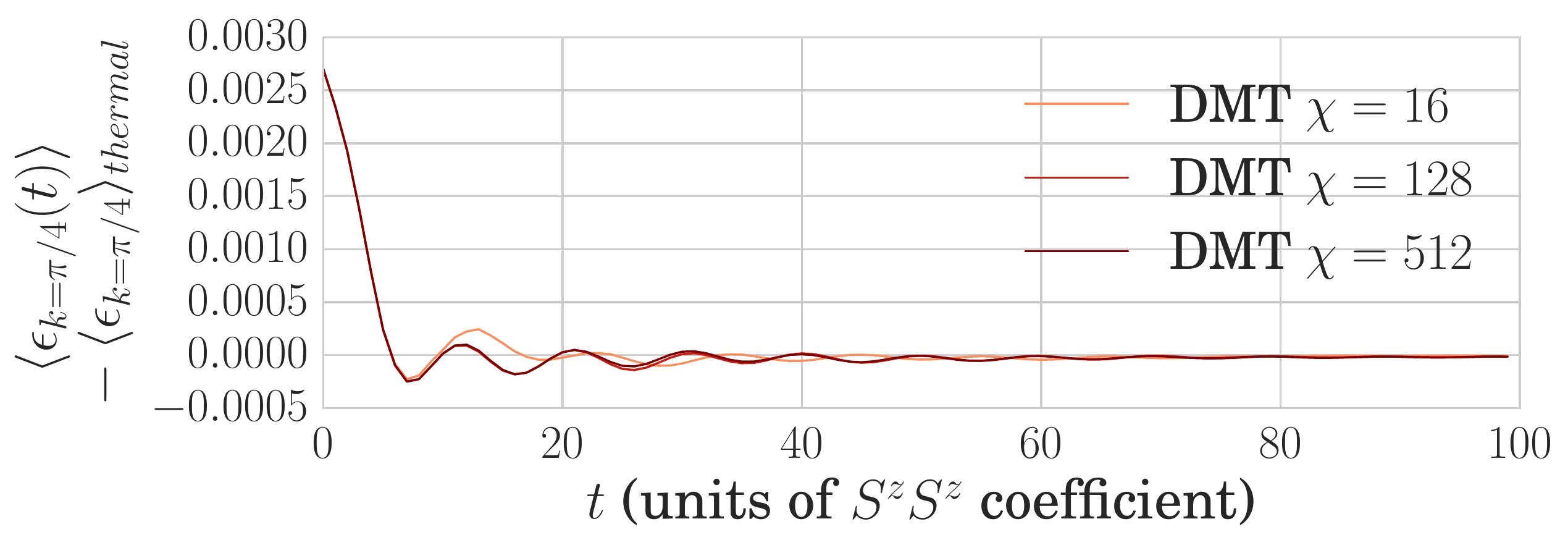}
		\\[-2ex]\hspace{-\textwidth}\begin{minipage}{0mm}\vspace{-52mm}\subfigure[]{\label{fig:MSffEE:epsfc:DMT}}\hspace{-27mm}\end{minipage}
	\end{minipage}
	\caption{Fourier component of energy density for purification time evolution and DMT starting from a far-from-equilibrium initial state on a 128-site chain.
	The thermal value is $\langle\epsilon_{k = \pi/4}\rangle_{\mathrm{thermal}} = -0.00031$.} 
	\label{fig:MSffEE:epsfc}
\end{figure}

\begin{figure}
	\begin{minipage}{83mm}
		\includegraphics[width=80mm]{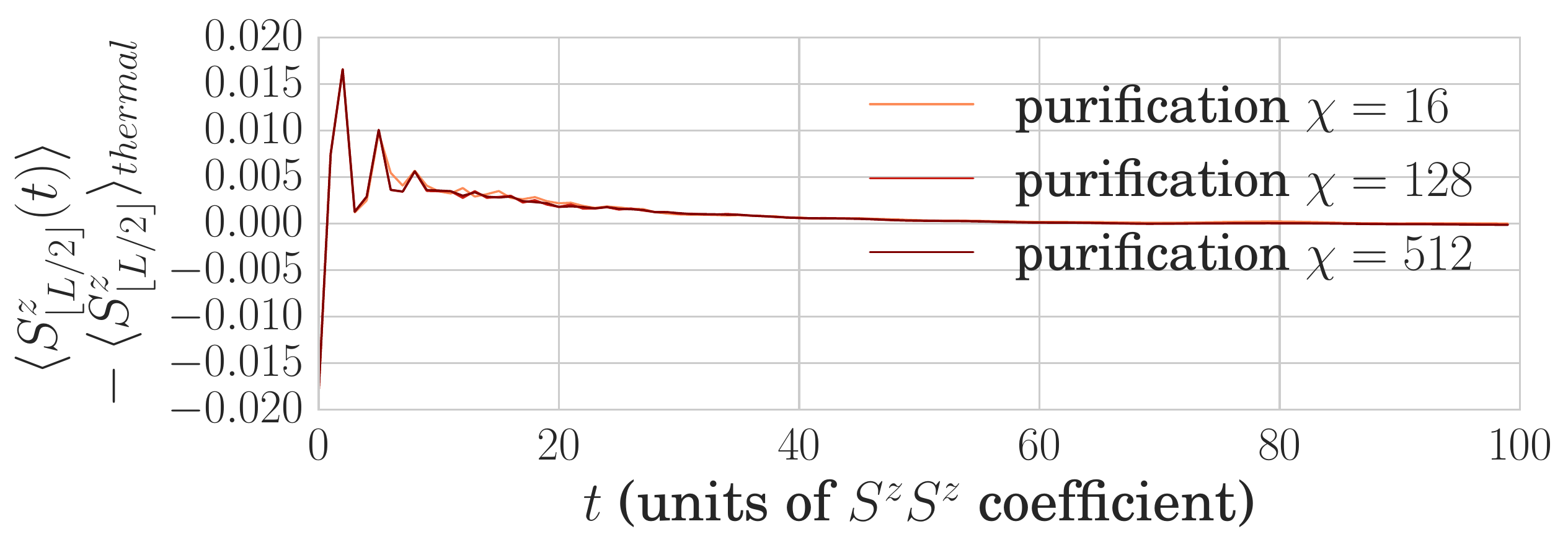}
		\\[-2ex]\hspace{-\textwidth}\begin{minipage}{0mm}\vspace{-52mm}\subfigure[]{\label{fig:MSffEE:Sz:KBM}}\hspace{-27mm}\end{minipage}
	\end{minipage}
	\begin{minipage}{83mm}
		\includegraphics[width=80mm]{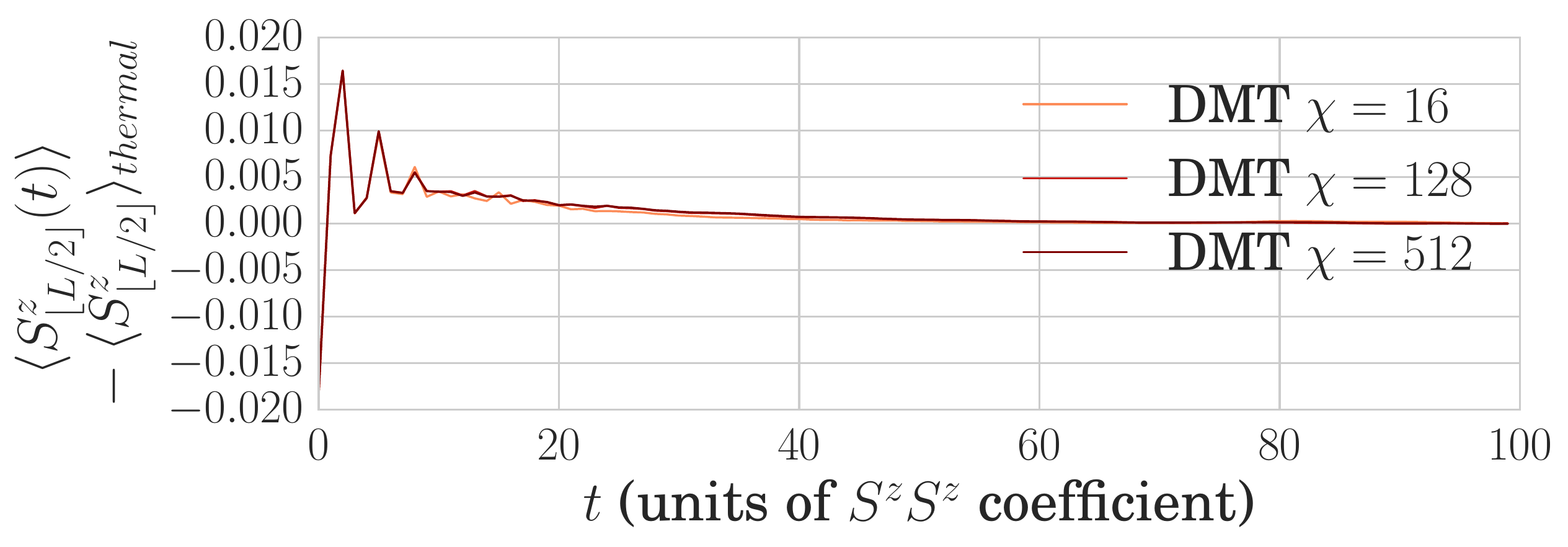}
		\\[-2ex]\hspace{-\textwidth}\begin{minipage}{0mm}\vspace{-52mm}\subfigure[]{\label{fig:MSffEE:Sz:DMT}}\hspace{-27mm}\end{minipage}
	\end{minipage}
	\caption{Expectation value of $S^z$ at the midpoint of the chain for purification time evolution and DMT starting from a far-from-equilibrium initial state on a 128-site chain.
	The thermal value is $\langle S^z_{\lfloor L/2 \rfloor} \rangle_{\mathrm{thermal}} = -0.021$.}
	\label{fig:MSffEE:Sz}
\end{figure}

In order to probe the performance of DMT for in more generic situations, we quench from a Gibbs state of the TFIM \eqref{eq:TFIM} with
\begin{align}
  h^x_0 &= 0.5\notag\\
  h^z_0 &= 0.5
\end{align}
and
\begin{equation}
  \beta_j = \beta_0 \big(1 + g'_j\big)
\end{equation}
where
\begin{equation}
  g'_j = 0.1 \times
  \begin{cases}
    0 & j \bmod 8 = \textrm{1, 2, 7, or 0}, \\
    1 & j \bmod 8 = \textrm{3, 4, 5, or 6}.
  \end{cases}
\end{equation}
(as in section \ref{SS:MSnEE}) to a TFIM \eqref{eq:TFIM} with
\begin{align}
  h^x_1 &= 2.0\notag\\
  h^z_1 &= 0.5.
\end{align}
We again compare to the purification method \cite{KBM2013}, and find that our method and that purification time evolution both converge quickly (see Figs. \ref{fig:MSffEE:epsfc} and \ref{fig:MSffEE:Sz}).
Even very small bond dimensions (e.g. $\chi = 16$) can accurately treat long-time, hydrodynamic behavior; accurately treating short-time behavior requires somewhat higher bond dimension.

\section{Convergence of mixed-state evolution}
\begin{figure*}
  \includegraphics[width=\textwidth]{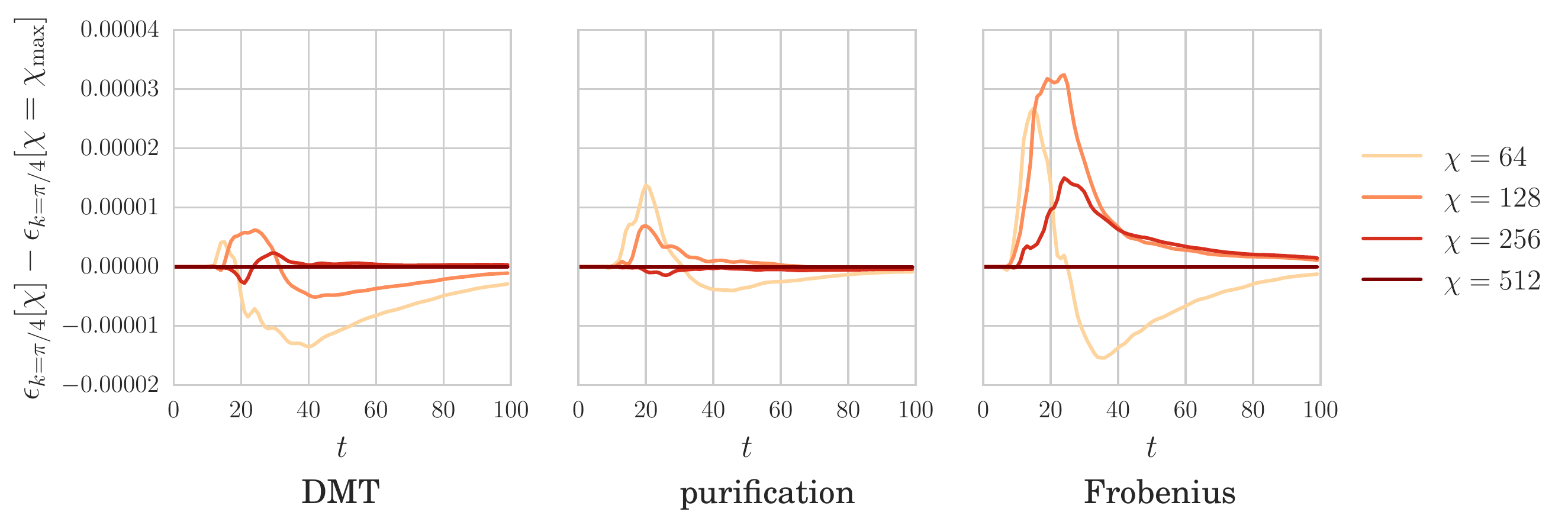}
  \caption{Convergence of $\epsilon_{k = \pi/4}$ for three algorithms.
    Initial state is a near-equilibrium mixed state (cf Section \ref{SS:MSnEE} and Figure \ref{fig:MSnEE:epsfc}) on a 128-site chain.
    For each algorithm, we plot $\epsilon_{k = \pi/4}[\chi] - \epsilon_{k = \pi/4}[\chi = \chi_{\mathrm{max}}]$---that is, how far the measurement during a run with a certain bond dimension $\chi$ deviates from measurement during a run with some high bond dimension.}
  \label{fig:APP:MSnEE:conv:epsfc}
\end{figure*}

\begin{figure*}
  \includegraphics[width=\textwidth]{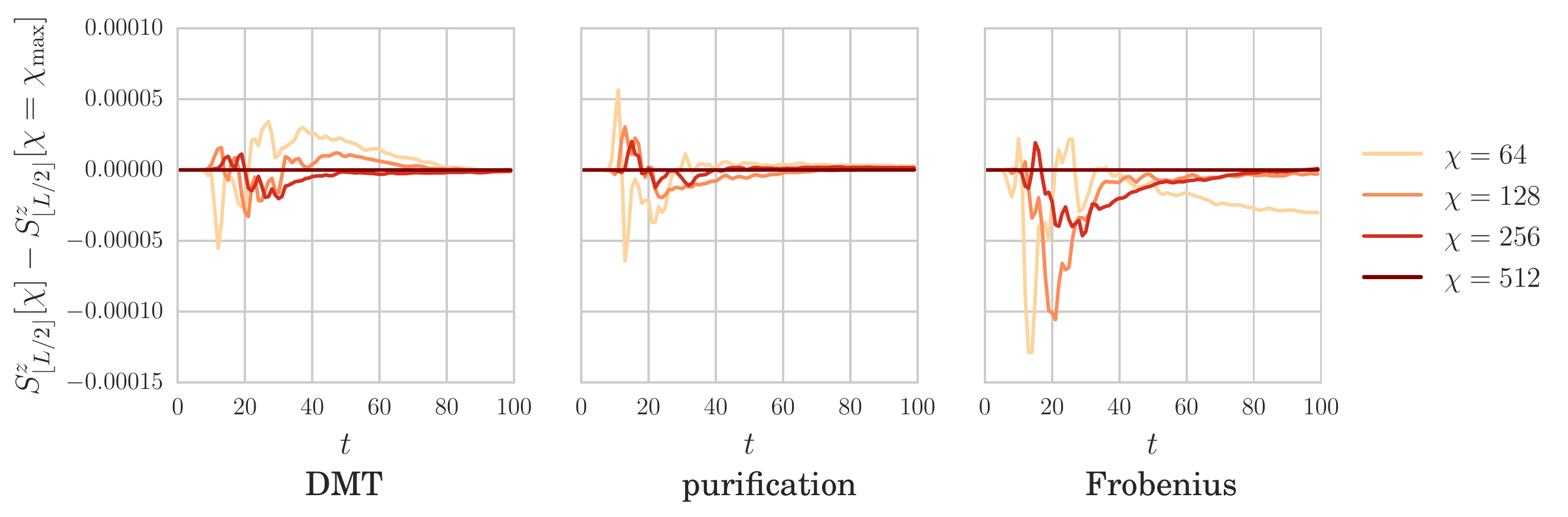}
  \caption{Convergence of $S^z$ at site $L/2$ for three algorithms.
    Initial state is a near-equilibrium mixed state (cf Section \ref{SS:MSnEE} and Figure \ref{fig:MSnEE:Sz}) on a 128-site chain.
    For each algorithm, we plot $S^z_{L/2}[\chi] - S^z_{L/2}[\chi = \chi_{\mathrm{max}}]$---that is, how far the measurement during a run with a certain bond dimension $\chi$ deviates from measurement during a run with some high bond dimension.}
    \label{fig:APP:MSnEE:conv:Sz}
\end{figure*}

\begin{figure*}
	\includegraphics[width=\textwidth]{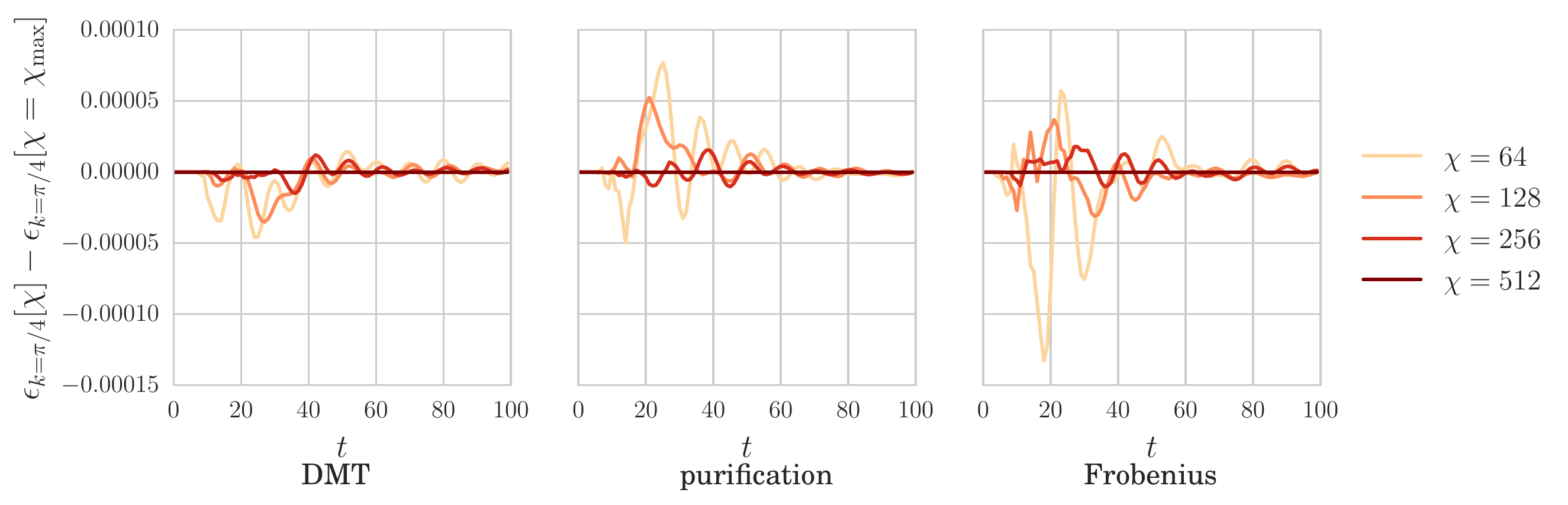}
	\caption{Convergence of $\epsilon_{k = \pi/4}$ for three algorithms. Initial state is a far-from-equilibrium mixed state (cf Section \ref{SS:MSffEE} and Figure \ref{fig:MSffEE:epsfc}) on a 128-site chain. For each algorithm, we plot $\epsilon_{k = \pi/4}[\chi] - \epsilon_{k = \pi/4}[\chi = \chi_{\mathrm{max}}]$---that is, how far the measurement during a run with a certain bond dimension $\chi$ deviates from measurement during a run with some high bond dimension.}
	\label{fig:APP:MSffEE:conv:epsfc}
\end{figure*}

\begin{figure*}
	\includegraphics[width=\textwidth]{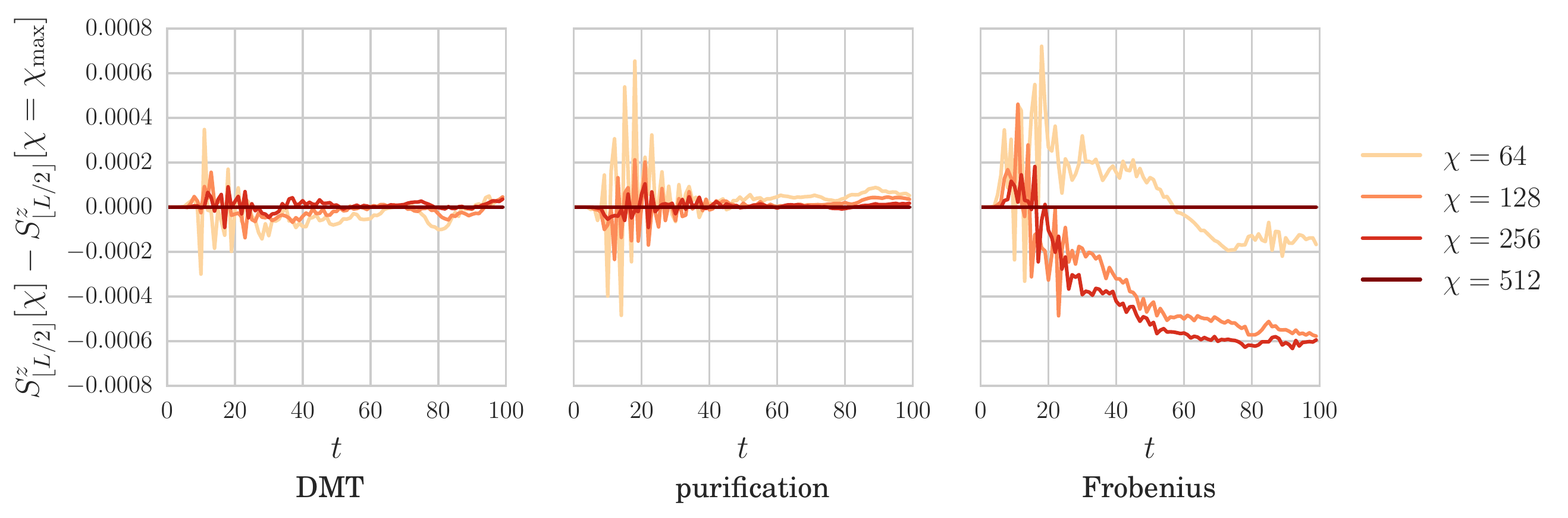}
	\caption{Convergence of $S^z$ at site $L/2$ for three algorithms. Initial state is a far-from-equilibrium mixed state (cf Section \ref{SS:MSffEE} and Figure \ref{fig:MSffEE:Sz}) on a 128-site chain. For each algorithm, we plot $S^z_{L/2}[\chi] - S^z_{L/2}[\chi = \chi_{\mathrm{max}}]$---that is, how far the measurement during a run with a certain bond dimension $\chi$ deviates from measurement during a run with some high bond dimension.}
	\label{fig:APP:MSffEE:conv:Sz}
\end{figure*}

It is difficult to judge convergence of any of the three algorithms from plots like Figure \ref{fig:MSnEE:epsfc} or \ref{fig:MSffEE:epsfc}.
In Figure \ref{fig:APP:MSnEE:conv:epsfc} we take a near-equilibrium initial state and plot the deviation in $\epsilon_{k = \pi/4}$, as measured for a series of bond dimensions $\chi$, from the last (largest) $\chi$ in the series.
In Figure \ref{fig:APP:MSffEE:conv:epsfc} we do the same for a far-from-equilibrium mixed state, and in Figures \ref{fig:APP:MSnEE:conv:Sz}, \ref{fig:APP:MSffEE:conv:Sz} for $S^z_{L/2}$ for near-equlibrium and far-from-equilibrium mixed states.

Our method converges with approximately the same bond dimension vs. accuracy tradeoff as purification time evolution for both the near-equilibrium initial state (Fig.~\ref{fig:APP:MSnEE:conv:epsfc}) and the far-from-equilibrium initial state (Fig.~\ref{fig:APP:MSffEE:conv:epsfc}).
In both cases, Frobenius time evolution converges more slowly than either method.

\bibliography{references.bib}
\end{document}